\documentclass[%
 reprint,
superscriptaddress,
nofootinbib,
 amsmath,amssymb,
 aps, prx
]{revtex4-2}

\usepackage{graphicx} 
\usepackage{dcolumn} 
\usepackage{bm} 
\usepackage[colorlinks=True,
    citecolor=black!70,
    urlcolor=blue!50!black,
    linkcolor=black!70
    ]{hyperref}
\usepackage{xcolor}

\usepackage{caption}
\usepackage{subcaption}

\usepackage{ragged2e} 
\DeclareCaptionJustification{justified}{\justifying}

\usepackage{amsthm}
\usepackage{physics}
\usepackage[capitalize,compress]{cleveref}

\crefname{section}{Section}{Section} 
\crefname{subsection}{Section}{Section}
\crefname{thm}{Theorem}{Theorems}
\Crefname{thm}{Theorem}{Theorems}

\crefname{corll}{Corollary}{Corollaries}
\theoremstyle{remark}

\theoremstyle{definition}

\crefname{equation}{Eq.}{Eqs.}
\Crefname{equation}{Equation}{Equations}
\usepackage{booktabs,multirow}

\newcommand{\N}{\mathcal N}
\newcommand{\T}{\mathcal T}

\newcommand{\Eu}{\mathbb{E}_{\mathbf{U}}}
\newcommand{\U}{\mathbf{U}}
\newcommand{\Ev}{\mathbb{E}_{\mathbf{V}}}
\newcommand{\V}{\mathbf{V}}

\newcommand{\x}{\mathbf{x}}

\newcommand{\bsigma}{\sigma}
\newcommand{\fsim}{\text{fSim}}
\newcommand{\pe}{\text{PE}}

\newcommand{\rhoi}{\rho_{\mathbf{U}}(0)}
\newcommand{\rhon}{\rho_{\mathbf{U}}(\N)}
\newcommand{\rhoN}{\rho_{\mathbf{U}}(\N)}

\newcommand{\rhotwo}{\overline{\rho^{(2)}}}
\newcommand{\com}[1]{{\color{olive}{[AD: #1]}}}

\DeclareMathOperator{\SWAP}{\mathcal{S}}
\DeclareMathOperator{\PROJ}{\mathcal{P}}

\DeclareMathOperator{\also}{\mathrm{ and }}

\usepackage{tikz}
\usetikzlibrary{quantikz}
\begin{document}

\preprint{APS/123-QED}

\title{A sharp phase transition in linear cross-entropy benchmarking}

\author{Brayden Ware}
\affiliation{Joint Quantum Institute, NIST/University of Maryland, College Park, MD 20742, USA}
\affiliation{Joint Center for Quantum Information and Computer Science, NIST/University of Maryland, College Park, MD 20742, USA}
\author{Abhinav Deshpande}
\thanks{Currently at IBM Quantum, Almaden Research Center, San Jose, California 95120, USA}
\affiliation{Institute for Quantum Information and Matter, California Institute of Technology, Pasadena, CA 91125, USA}
\author{Dominik Hangleiter}
\affiliation{Joint Center for Quantum Information and Computer Science, NIST/University of Maryland, College Park, MD 20742, USA}
\author{Pradeep Niroula}
\affiliation{Joint Quantum Institute, NIST/University of Maryland, College Park, MD 20742, USA}
\affiliation{Joint Center for Quantum Information and Computer Science, NIST/University of Maryland, College Park, MD 20742, USA}
\author{Bill Fefferman}
\affiliation{Department of Computer Science, University of Chicago, Chicago, Illinois 60637, USA}
\author{Alexey V.\ Gorshkov}
\affiliation{Joint Quantum Institute, NIST/University of Maryland, College Park, MD 20742, USA}
\affiliation{Joint Center for Quantum Information and Computer Science, NIST/University of Maryland, College Park, MD 20742, USA}
\author{Michael J. Gullans}
\affiliation{Joint Center for Quantum Information and Computer Science, NIST/University of Maryland, College Park, MD 20742, USA}
\date{\today}

\begin{abstract}
Demonstrations of quantum computational advantage and benchmarks of quantum processors via quantum random circuit sampling are based on evaluating the linear cross-entropy benchmark (XEB). 
A key question in the theory of XEB is whether it approximates the fidelity of the quantum state preparation. 
Previous works have shown that the XEB generically approximates the fidelity in a regime where the noise rate per qudit $\varepsilon$ satisfies  $\varepsilon N \ll 1$ for a system of $N$ qudits
and that this approximation breaks down at large noise rates.
Here, we show that the breakdown of XEB as a fidelity proxy occurs as a sharp phase transition at a critical value of $\varepsilon N$ that depends on the circuit architecture and properties of the two-qubit gates, including in particular their entangling power. 
We study the phase transition using a mapping of average two-copy quantities to statistical mechanics models in random quantum circuit architectures with full or one-dimensional connectivity. 
We explain the phase transition behavior in terms of spectral properties of the transfer matrix of the statistical mechanics model and identify two-qubit gate sets that exhibit the largest noise robustness.
\end{abstract}

\maketitle


\section{Introduction \label{sec:intro}}

Near-term quantum devices consist of dozens of imperfect qubits, with error rates far too large to outperform classical computers in tasks like factoring large integers.
Instead, several recent attempts to demonstrate the computational power of noisy quantum devices have focused on \textit{quantum random circuit sampling (RCS)} \cite{boixo_characterizing_2018,aruteQuantumSupremacyUsing2019,wu_strong_2021,zhu_quantum_2022,hangleiter_computational_2023}. 
In RCS, the task is to repeatedly produce and measure a single, highly entangled many-body wavefunction $\psi(x)$, generating a sequence of outcomes $\{x_i\}$ that ideally occur with probabilities $p(x) = | \psi(x)|^2$. 
As generating complex entangled wavefunctions is a core capability of quantum computers that cannot be replicated by classical computers (see e.g., \cite{bremner_average-case_2016,boixo_characterizing_2016,bouland_complexity_2019,movassagh_quantum_2020}), RCS is a natural task for testing prototype quantum devices.

Of course, error prone qubits do not allow for precisely reproducing the same wavefunction repeatedly, or for precisely sampling from the correct distribution $p(x)$. 
To compare the performance of a particular quantum device with disparate quantum devices or classical computers at this task, benchmarks---quantitative measures of success---are needed. 
A straightforward but strict benchmark is the fidelity between the target wavefunction and the output wavefunction.
However, measuring fidelity in large quantum systems is intractable on current hardware since the best available schemes require the application of deep noise-free random Clifford circuits for shadow fidelity estimation \cite{huang_efficient_2021,kliesch_theory_2021}. 
Moreover, when comparing with classical algorithms, the fidelity is only meaningful for certain algorithms that also produce a classical representation of the wave function.  

In place of fidelity, the \emph{linear cross-entropy benchmark} (XEB) has been used as a measure of success in RCS \cite{boixo_characterizing_2018,aruteQuantumSupremacyUsing2019,dalzellRandomQuantumCircuits2021a,gao_limitations_2021} and beyond \cite{choiEmergentQuantumRandomness2023,mark_benchmarking_2022, liCrossEntropyBenchmark2022a}. 
The XEB  
\begin{equation}
    \chi = 2^N \sum_x \overline{p(x) q(x)} - 1
\end{equation}
measures the correlation between the ideal distribution $p(x)$ and the actual distribution $q(x)$ of sample outcomes, averaged over the random circuit instances. 
Importantly, we expect that the XEB can be
sample-efficiently estimated just from the produced samples $x_i$ of a few random quantum circuits.\footnote{See \cite[Sec. V.B.3]{hangleiter_computational_2023} for the argument.}
This sample efficiency is crucial for the usefulness of the benchmark. However, the ideal probabilities $p(x)$ of the measured samples must still be known to estimate the benchmark, and these must be computed via classical algorithms with exponential runtime.

There are two crucial theoretical questions about XEB for understanding how well it serves its purpose as a  benchmark for RCS. 
First, one must understand what scores are, in principle, achievable by noisy quantum devices. 
Secondly, to use XEB as a benchmark of computational power, one must understand what scores are achievable by competing classical devices, running arbitrary algorithms. 

\begin{figure}
    \includegraphics{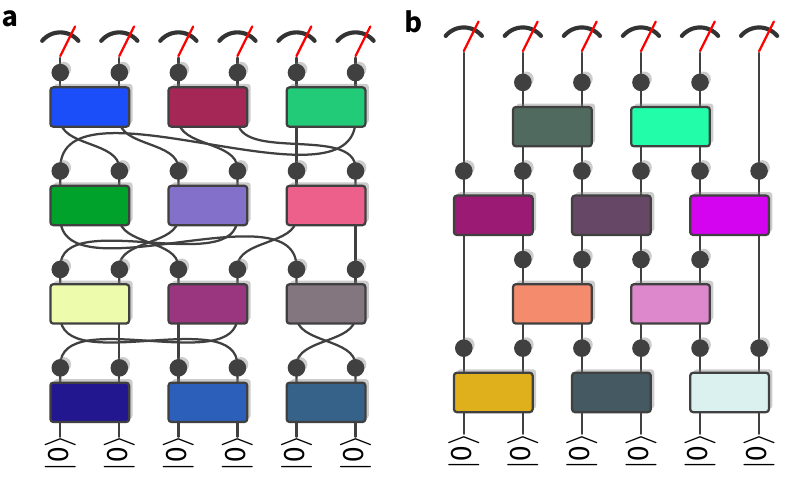}
    \caption{Circuit geometries considered in this paper. The evolution consists of layers of random two-qudit gates chosen either from a Haar random distribution or according to \cref{eq:gate} and single-qudit noise (shown as black circles) specified by an arbitrary (but fixed) noise channel $\mathcal{N}$ applied after each gate layer. (a) In the all-to-all geometry, the qubits are permuted randomly between gate layers. (b) In the 1D geometry, alternating layers of gates are applied in a brickwork fashion with open boundary conditions.}
    \label{fig:cartoon}
\end{figure}

In this work, we address the first question by phrasing it as a comparison between XEB and fidelity: can high XEB scores be achieved without producing states with high fidelity to the target? This question has been studied previously in work of \textcite{gaoLimitationsLinearCrossEntropy2021,dalzellRandomQuantumCircuits2021a}, who come to the conclusion that XEB closely approximates fidelity in a regime of significantly weak, spatially homogenous, and temporally independent noise. For an $N$-qubit system with local noise rate $\varepsilon$, this regime is characterized by $\varepsilon N \ll 1$. 
However, they have also shown that violations of these assumptions break the correlation between XEB and fidelity. 
Our work addresses the exact dependence of the match between XEB and fidelity as the noise is increased. We show that this match breaks suddenly as noise is increased, occurring as a sharp phase transition in the behavior of XEB at a critical value of $\varepsilon N $. 
This remains true for variants of the XEB \cite{dalzellRandomQuantumCircuits2021a} which, in the low-noise limit, track the fidelity in a wider parameter regime. 
In addition to explaining the mechanism behind the transition, we investigate how the critical noise value depends on the gate set and circuit geometry used.

In order to analyze the noise dependence of the XEB, we make use of a mapping between average properties of two copies of the quantum wavefunction---which includes both the fidelity and the XEB---to observables of a statistical model \cite{zhou_emergent_2019,vasseurEntanglementTransitionsHolographic2019,hunter-jones_unitary_2019,barak_spoofing_2021,dalzellRandomQuantumCircuits2021a,gaoLimitationsLinearCrossEntropy2021, liEntanglementDynamicsNoisy2023,deshpande_tight_2022}.
In this model, the average dynamics of the quantum system can be analyzed in terms of a transfer matrix on a $2^N$-dimensional configuration space, and the behavior of the XEB can be explained via spectral properties of this transfer matrix.
We analyze the model using simulations in two extremal circuit architectures, namely a fully connected (all-to-all) architecture, and a one-dimensional architecture (see \cref{fig:cartoon}), and find the same qualitative behavior. This suggests that the phase transition in the XEB is in fact a universal feature of random quantum circuits in any architecture.

In the all-to-all model we exploit the permutation symmetry of the qudits to exactly solve the corresponding statistical mechanical model with large numbers of qudits in terms of an exponentially reduced state space. 
To analyze one-dimensional circuits, we use matrix-product state techniques for computing the dynamics and spectra of the statistical model.

In the following, we will first introduce the statistical mechanical model for the two architectures (\cref{sec:background}), then show our main results on the sharp phase transition of the XEB in \cref{sec:phase transition} for the paradigmatic gate set of Haar-random gates, and finally discuss the gate-set dependence of the location of the transition in \cref{sec:gatedependence}.

\section{\label{sec:background} Background}

\subsection{\label{sec:setup} Random circuit observables}

We study noisy random circuits acting on $N$ qudits of dimension $q$. The circuits consist of two-qudit gates generated in one of two ways: Haar-random gates $G$ drawn from $U(q^2)$, or by using a specific two-qudit gate $g$ and Haar-random single-qudit gates $U_1, U_2, U_3, U_4 \in U(q)$:
\begin{equation}
G = (U_1 \otimes U_2) g \ (U_3 \otimes U_4).
\label{eq:gate}
\end{equation}

The circuits we study (illustrated in \cref{fig:cartoon}) consist of $d$ layers of $N/2$ gates, each of which connects either 
random disjoint pairs of qudits (the `all-to-all architecture', see  \cref{sec:all}), or neighboring qudits of a chain, alternating in a brickwork pattern (the 1D architecture, see \cref{sec:1d}).
After each layer of gates, each qudit experiences noise as specified by an arbitrary single-qudit noise channel $\N$.
We initialize the circuit with the product state
\((\ket{0}\bra{0})^{\otimes N}\).

Our reference quantity is the average fidelity $F = \Eu \Tr[\rhon \rhoi]$ between the noisy output state $\rhon$ and the corresponding noiseless pure state $\rhoi$, where the average is taken over the random circuit instance $\mathbf U$. 
The average fidelity $F$ takes the value $1$ in the ideal limit of no noise, and the value $q^{-N}$ when $\rhon$ and $\rhoi$ are totally uncorrelated.
%
A convenient reformulation of the fidelity, 
\begin{equation} \label{eq:def-F}
    F = \Tr(\SWAP^{\otimes N}  \Eu \left[ \rhoi \otimes \rhon \right]),
\end{equation} 
considers two copies of the same circuit, one of which is evolved with a noisy evolution and the
other with the noiseless evolution, followed by the $\SWAP^{\otimes N}$ operator that swaps corresponding qudits between the two copies.
The second important quantity we will consider in this paper is the \textit{linear cross entropy benchmark} (XEB)
\begin{align}
    \chi &= q^N \mathbb{E}_{\mathbf{U}}\sum_\x \bra{\x}\rhon \ket{\x} \bra{\x} \rhoi \ket{\x} - 1 \nonumber
    \label{eq:def-X}
\end{align}
which has been proposed as a classical estimator of the average fidelity for random quantum circuits. 
It can be written as a two-copy expectation value in a similar way:
\begin{equation}
    \chi =   \Tr\left( \left[q^N \PROJ^{\otimes N}  -1\right] \Eu \left[\rhon \otimes \rhoi\right] \right), 
\end{equation}
where \begin{equation}
\PROJ = \sum_{x=0}^{q-1} \ket{x}\bra{x} \otimes \ket{x}\bra{x}.
\end{equation}

Since the values of $\chi$ for low-depth random circuits can be much larger than one, the variant quantity $\chi_B = \chi/(q^N Z - 1)$ has been proposed by \textcite{dalzellRandomQuantumCircuits2021a} as a better estimator of the fidelity\footnote{See also Refs.~\cite{rinott_statistical_2022,choiEmergentQuantumRandomness2023,mark_benchmarking_2022,liu_benchmarking_2022} for similar suggestions.}.
Here, 
\begin{align}
    Z &= \mathbb{E}_{\mathbf{U}}\sum_\x \bra{\x}\rhoi \ket{\x} \bra{\x} \rhoi \ket{\x} \nonumber \\
    &=  \Tr(\PROJ^{\otimes N} \Eu \left[\rhoi \otimes \rhoi \right]),
    \label{eq:def-Z}
\end{align}
is the collision probability of the noiseless circuit.
Like the fidelity, $\chi_B$ takes values between $0$ and $1$, and attains its maximal value in the noiseless case. 
As we will see below, which precise variant of XEB is used will not affect the behavior of the phase transition.

We now introduce the statistical model \cite{dalzellRandomQuantumCircuits2021a} used to analyze the noise-dependence of these two-copy average quantities. 

\subsection{\label{sec:stat}Statistical model}

For an arbitrary pair of density matrices $\rho_A, \rho_B$ on $N$ qudits of dimension $q$, consider the Haar-averaged two-copy density matrix
\begin{equation}\rhotwo = \Ev (\V \rho_A \V^{\dagger}) \otimes (\V \rho_B \V^{\dagger}),\end{equation}
where $\V$ is a product of single-qudit Haar random rotations
\begin{equation}
\V = \prod_{i=1}^N V_i.
\end{equation}
Such an average has the result of projecting arbitrary two-copy density matrices $\rho_A \otimes \rho_B$ into the symmetric subspace, i.e., the space invariant under arbitrary rotations $\mathbf V \otimes \mathbf V$. 
The symmetric subspace is spanned by tensor products of \(I/q^2\) and
\(\SWAP/q\), where $I$ is the two-qubit identity operator and $\SWAP$ is the operator that swaps two corresponding qudits in the two copies. 
Thus, the average can be expressed as
\begin{equation}
\rhotwo = \sum_{\bsigma \in \{0, 1\}^{\otimes N}} p(\bsigma) \bigotimes_{i=1}^N \left( \frac{I}{q^2} \right)^{1-\sigma_i} \left( \frac{\SWAP}{q}\right)^{\sigma_i}.
\label{eq:rho2}
\end{equation}
We reinterpret $\rhotwo$ as a vector $\ket{\rho}$ in a many-body state space, using basis states $\ket{\sigma}$ with \(\bsigma \in \{0, 1\}^{\otimes N}\) corresponding to the terms of the above sum, so that
\begin{equation}
    \ket{\rho} = \sum_\sigma p(\sigma) \ket{\sigma}. 
\end{equation}
Using this representation, the fidelity of $\rho_B$ with respect to $\rho_A$ and their linear cross-entropy can be expressed as inner products with fixed vectors: $F = \Tr(\mathbb{S} \rhotwo) = \braket{\mathbb{S}}{\rho}$ and $X=\braket{\mathbb{P}}{\rho}$, where $\mathbb{S} = \mathcal{S}^{\otimes N}$ and $\mathbb{P} = \mathcal{P}^{\otimes N}$. Similarly, the trace of $\rho_A \otimes \rho_B$ is $\braket{\mathbb{I}}{\rho} = 1$. These vectors are expressed as follows, using the dual basis $\bra{\sigma}$ defined via $\braket{\sigma }{ \sigma'} = \delta_{\sigma \sigma'}$:
\begin{gather*} \label{eq:observables}
\bra{\mathbb{I}} = (\bra{0} + \bra{1})^{\otimes N}, \,
\bra{\mathbb{P}} = (q^{-1} \bra{0} + \bra{1})^{\otimes N}, \, \\
\also \bra{\mathbb{S}} = (q^{-1} \bra{0} + q\bra{1})^{\otimes N}.
\end{gather*}

In this paper, we compute the circuit-averaged fidelity $F$ and the XEB $\chi= q^N X - 1$ by computing the two-copy density matrix 
\begin{equation} \rhotwo = \Eu \left[\rho_{\U}(0) \otimes \rho_{\U}(\N)\right]. \end{equation}
If the gates are randomly chosen from a distribution that is invariant under single qudit rotations --- such as the Haar distribution or the distribution in \cref{eq:gate} --- the averaged two-copy density matrix after each layer of gates has the form of Eq.\ (\ref{eq:rho2}).
As the gates are drawn independently at random between layers, 
$p(\bsigma)$ after a given layer depends only on the distribution $p$ from the previous layer.
The corresponding evolution of $p$ under the circuit layer $l$ is thus governed by a transfer matrix $\T_l(\bsigma', \bsigma)$. 
This transfer matrix can be constructed by composing a matrix $M$ corresponding to each random two-qudit gate and a matrix $N$ for each  single qudit noise channel as follows:
\begin{equation} \label{eq:transfermatrixlayer}
\T_l(\bsigma', \bsigma) = \prod_i N_{\sigma'_i , \sigma_i} \prod_{\langle ij \rangle} M_{\sigma'_i \sigma'_j, \sigma_i \sigma_j},
\end{equation}
where $\langle ij \rangle$ denotes pairs of qudits $i, j$ which are acted upon by a random gate in a particular layer $l$ of the circuit. 
As $M$ denotes the update corresponding to the average over a unitary gate, its form is constrained by the requirement that it preserves the trace and the  fidelity. 
We show in \cref{app:deriv} that the two-qudit update matrix $M$ must take the following form\footnote{Note that an equivalent but distinct parameterization of the update rule was described in Ref.~\cite{gaoLimitationsLinearCrossEntropy2021}---translations between our conventions and those of Ref.~\cite{gaoLimitationsLinearCrossEntropy2021} are given in Appendix~\ref{app:deriv}.}
\begin{equation} \label{eq:M}
    M = 
    \begin{bmatrix}
        1 &  \frac{\alpha q^2}{q^2+1} &  \frac{\alpha q^2}{q^2+1} & 0\\
        0 & 1-\alpha - \beta & \beta & 0\\
        0 & \beta & 1-\alpha-\beta & 0\\
        0 & \frac{\alpha}{q^2+1} &  \frac{\alpha}{q^2+1} & 1
    \end{bmatrix},
\end{equation}
where we order the two-site $\{I/q^2, \SWAP/q\}$ basis (corresponding to $\sigma \in \{0,1\}$) of the statistical model as $\sigma_i\sigma_j \in \{ 00, 01, 10, 11 \}$ and $\alpha, \beta \in \mathbb R$. 
This form follows from the conservation of the trace and fidelity under noiseless operations, as well as an additional principle that guarantees reflection symmetry for $M$. 
For Haar-random two-qudit gates, $\alpha = 1$ and $\beta = 0$ \cite{dalzellRandomQuantumCircuits2022}; the computation of $\alpha$ and $\beta$ corresponding to other gate sets is detailed in Appendix~\ref{app:deriv}.
The parameter $\alpha$ can be interpreted as controlling the rate at which $\SWAP$ factors, corresponding to $\sigma = 1$, are duplicated ($I \otimes \SWAP \to \SWAP \otimes \SWAP$) and destroyed ($I \otimes \SWAP \to I \otimes I$). The parameter $\beta$ on the other hand controls the rate at which $\SWAP$ factors hop between sites $I \otimes \SWAP \to \SWAP \otimes I$.

Similarly, the update rule $N$ for a single-qudit noise channel $\N$ can be expressed using the $\{I/q^2, \SWAP/q\}$ basis as \cite{dalzellRandomQuantumCircuits2021a}
\begin{equation}
N = 
    \begin{bmatrix}
        1  & \gamma \\
        0 & 1-\gamma
    \end{bmatrix}.
\end{equation}
The parameter $\gamma$ sets a rate at which $\SWAP$ factors decay $\SWAP \to I$.
It is shown in Appendix~\ref{app:deriv} that $\gamma$ is equal to the infidelity of $\N$ with respect to the identity channel, averaged over pure state inputs, and rescaled by $q/(q-1)$.

The corresponding rate for $I$ factors to decay to $\SWAP$ is zero, as the maximally mixed two-copy density matrix is a fixed point of $\N$. We point out that, for this to be true, no conditions are required on the noise channel $\N$---in particular, we do not need to assume it to be unital. While the action of a non-unital noise channel $\N$ (on one or both copies) will not leave the maximally mixed density matrix $I \otimes I$ fixed, when $\N$ acts on only one copy the following average over single-qudit Haar rotations projects $\N(I) \otimes I$ back onto $I \otimes I$. When instead noise acts on both copies of the two-copy density matrix, as occurs when computing the purity or the collision probability of the noisy density matrix, only unital noise channels preserve the maximally mixed density matrix $I \otimes I$. We expand upon this in Appendix~\ref{sec:purity}.

In summary, the statistical models we consider are defined through a transition matrix $\T(\alpha, \beta, \gamma) $, parameterized by three parameters: $\alpha$ and $\beta$, which are derived as a property of the gate set, and $\gamma$, which is proportional to the infidelity of the noise channel.
The case $\alpha=1$ and $\beta=0$ corresponds to Haar-random two-qudit gates; we will consider this case primarily (see \cref{sec:phase transition}), and then show that other values of $\alpha$ and $\beta$ give rise to similar behavior (see \cref{sec:gatedependence}).

Let us now discuss the concrete implementation of the transfer matrix $\T$ for the all-to-all geometry (\cref{sec:all}) and the one-dimensional geometry (\cref{sec:1d}). 

\subsection{\label{sec:all}All-to-all circuit}
We first consider an all-to-all geometry, which is given by a fully connected architecture on which gates act on random pairs of qudits---without regard to spatial locality---in dense layers that cover all of the qudits. 
In addition to averaging over the randomness of the gates, we also average over all possible choices of the pairings of qudits; or equivalently, as depicted in \cref{fig:cartoon}(a), after each layer of gates the qudits are randomly permuted. 
This geometry allows for an efficient solution of the statistical model that exploits the permutation symmetry of the circuit architecture. 
A similar random circuit architecture, considered for example in Ref.~\cite{dalzellRandomQuantumCircuits2022}, has the gates acting on one pair of qudits at a time rather than in dense layers. We expect that our results would translate to this geometry with minimal changes.

We first note that the transfer matrix in the all-to-all circuit does not depend on the specific gate layer since all layers are constructed identically. It is constructed by averaging over all permutations of the qudits as
\begin{equation}
\T(\bsigma', \bsigma) = \frac{1}{N!}\sum_{P\in S_N} \T(\bsigma', P \bsigma),
\end{equation}
where $S_N$ denotes the permutation group of $N$ elements. 
By construction, the transfer matrix $\T(\bsigma', \bsigma)$ only depends on the Hamming weights of $\bsigma$ and $\bsigma'$, that is the number of sites with $\sigma_i=1$. We denote the Hamming weight of the bitstring $\bsigma$ as \(|\bsigma|\). $|\bsigma|$ just counts the number of tensor factors that are \(\SWAP/q\) in the corresponding term of $\rhotwo$. 

We denote the total probability of all configurations with a fixed $|\bsigma|=S$ as 
\begin{equation}
p_S = \sum_{|\bsigma|=S} p(\bsigma).
\end{equation} 
Since we are averaging over a permutation-symmetric family of circuits, we can restrict our evolution of $p(\bsigma)$ to an evolution of $p_S$, with a transfer matrix $\T_{S',S}.$ As $S$ ranges from $0$ to $N$, this is an $(N+1)\times(N+1)$ sized matrix, which is easy to study numerically for large system sizes.

Furthermore, an explicit formula for $\T_{S', S}$ can be written down with a bit of combinatorial logic, avoiding the need to explicitly compute the average over permutations. We assume for simplicity that $N$ is even, so that each qudit encounters a gate in each layer. First, we note that we can construct $\T_{S', S}(\alpha, \beta, \gamma)$ by composing a transfer matrix $M_{S', S}(\alpha, \beta)$ for the permutation-averaged action of the two-qudit gates with the action of a layer of noise channels $N_{S', S}(\gamma)$:
\begin{equation}
\T_{S', S}(\alpha, \beta, \gamma) = N_{S', S}(\gamma) M_{S', S}(\alpha, \beta).
\end{equation}
Secondly, we note that $M_{S', S}(\alpha, \beta)$ in fact does not depend on $\beta$, because the only effect of $\beta$ is to connect bitstrings $\bsigma$ which are the same up to permutations. Finally, we construct the permutation-averaged $M$ by explicitly counting the number of permutations which yield each possible value of $M(\bsigma', \bsigma)$: the result is
\begin{gather}
M_{S', S} = \frac{1}{\binom{N}{S}} \sum_{n_0, n_{1}, n_{2}}  2^{n_1} \binom{N/2}{n_0, n_{1}, n_{2}} \delta_{S, n_1 + 2n_2} \left[\vphantom{\sum_{x} x}\right. \nonumber \\
\left. \sum_{a,b,c} \binom{n_1}{a,b,c}
\left( \frac{q^2 \alpha}{q^2+1} \right)^{a}(1-\alpha)^{b}\left( \frac{\alpha}{q^2+1} \right)^{c}\delta_{S', b + 2c + 2n_2} 
\right].
\end{gather}
In this sum, $n_0, n_1, n_2$ specify the number of gates which encounter, respectively, $0$, $1$, or $2$ sites with $\sigma_i = 1$. Of the $n_1$ gates that encounter $1$ such site, 
$a, b, c$ are the number of gates which output $0$, $1$, or $2$ sites with $\sigma'_i=1$, respectively.

For $N(\bsigma', \bsigma)$, we see that the matrix elements only depend on $\bsigma$ and $\bsigma'$ through their Hamming weights $S$, $S'$, and so averaging over permutations yields
\begin{gather}
N_{S', S} = \binom{S}{S'} (1-\gamma)^{S'} \gamma^{S-S'}.
\end{gather}
In words, in a layer of noise, each of the $S$ factors of $\SWAP$ that appear in the input has a probability $\gamma$ to decay.

The numerical implementation of $\T_{S', S}$ is straightforward. However, we find it necessary to use high precision numerics to get accurate answers, particularly when $N$ is increased. Throughout, the all-to-all geometry results use BigFloats with $256$ bits of precision, as implemented in Julia \cite{bezanson2017julia}.

\subsection{\label{sec:1d} One-dimensional circuit}

Next we consider a one-dimensional brickwork-circuit architecture where all gates are nearest-neighbor, as depicted in \cref{fig:cartoon}(b). We use open boundary conditions throughout, and only consider circuits with an even number of qudits $N$; as such, in every other layer one site on each edge experiences neither a gate nor noise.

For our statistical model simulations, we use matrix-product states (MPS) to represent $\ket{\rho}$. We use the time-evolving block decimation (TEBD) algorithm \cite{schollwockDensitymatrixRenormalizationGroup2011a, vidalEfficientClassicalSimulation2003, vidalEfficientSimulationOneDimensional2004} to apply the transfer matrix $\T$ as a sequence of gates, as in \cref{eq:transfermatrixlayer}. Truncations of small singular values are used to keep the bond dimension of the MPS from growing uncontrollably. All simulation results are checked pointwise for convergence in the total truncated weight $\epsilon$, with truncation errors as small as $\epsilon=10^{-50}$ used in some parameter regimes to reach convergence. This is achieved by using BigFloats with $256$ bits of precision for calculations requiring $\epsilon \leq 10^{-30}$.

In addition to TEBD, we use the global MPS-Krylov algorithm \cite{paeckelTimeevolutionMethodsMatrixproduct2019a} to compute several leading eigenvalues of the transfer matrix. This algorithm is identical to the commonly used Arnoldi method \cite{arnoldiPrincipleMinimizedIterations1951} with MPS representations of all Krylov vectors. 

\begin{figure*}[htb] 
     \centering
         \includegraphics[width=\textwidth]{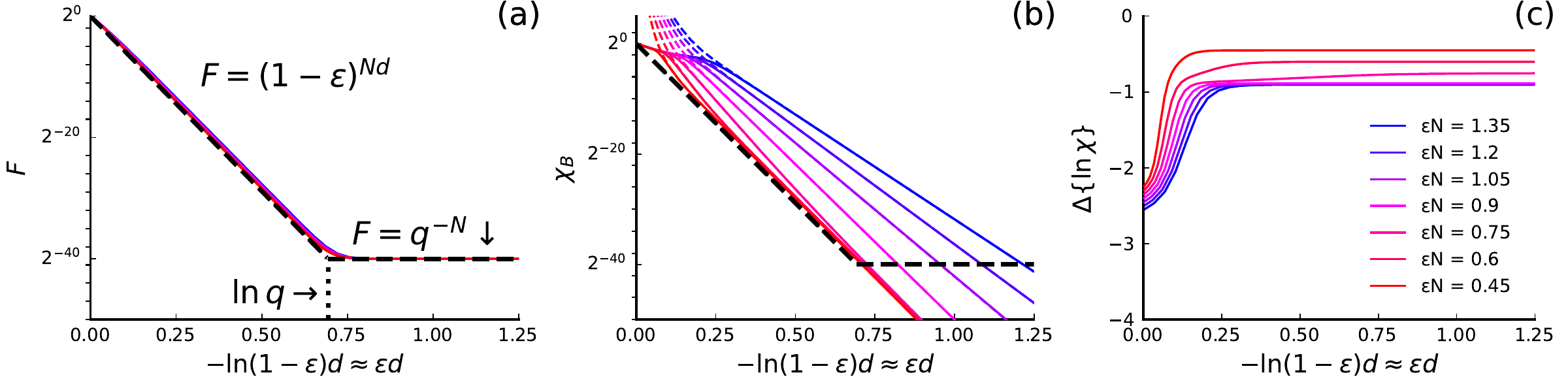}
   \caption{ \raggedright
   Fidelity and XEB in the all-to-all circuit model with $N=40$ qubits and Haar-random gates, for various values of the noise rate $\varepsilon$ (see color code in (c)), plotted against rescaled circuit depth.
   (a) Fidelity $F$ curves collapse onto the global-white noise prediction $(1-\varepsilon)^{Nd}$ until saturating at $q^{-N}$.
   (b) Normalized XEB $\chi_B$ as a function of rescaled circuit depth (solid lines) and unnormalized XEB $\chi$ (dashed lines), which are essentially equivalent after an initial equilibration time. For low noise $\varepsilon N \lesssim 0.92$, the XEB decay rate matches that of fidelity.
   (c) Decay rate of the XEB. For $\varepsilon N \gtrsim 0.92$, the decay rate saturates to a $\varepsilon N$-independent constant, $\Delta \{\ln \chi \}\approx -0.92$.
   \label{fig:Fvsx}}
\end{figure*}

\subsection{\label{sec:noiseless} Review of noiseless circuit results}

We now review some of the important properties of the statistical model considered in this paper that have been derived in previous works.
The statistical model has two fixed points in the noiseless ($\gamma=0$) limit; the important features of the dynamics can be understood in relation to these fixed points \cite{dalzellRandomQuantumCircuits2022, dalzellRandomQuantumCircuits2021a}. These fixed points are the dominant right-eigenvectors (with eigenvalue $1$) of the transfer matrix $\T$, and they are represented in our notation as \begin{align}
\ket{\mathbb{I}}\equiv \ket{0 \cdots 0} \, \also \, \ket{\mathbb{S}} \equiv \ket{1 \cdots 1}.
\end{align}
These eigenvectors represent two-copy density matrices proportional to the maximally mixed density matrix (proportional to $\mathbb{I}$) and the global swap operator $\mathbb{S}$, respectively. 
The corresponding left eigenvectors of $\T$ are $\bra{\mathbb{I}}$ and $\bra{\mathbb{S}}$, respectively; these left eigenvectors directly manifest the fact that the noiseless evolution preserves the 
trace and fidelity, recalling \cref{sec:stat}.

Without noise, all sufficiently deep circuits converge to a linear combination of $\ket{\mathbb{I}}$ and $\ket{\mathbb{S}}$. The specific linear combination is in fact fully determined by the two conserved quantities, trace and fidelity. When the evolution is initialized with a pure state the initial fidelity is given by $F=1$. 
This constrains the fixed-point state to be 
\begin{equation}
\ket{H} = \frac{q^N}{q^N+1}\ket{\mathbb{I}} + \frac{1}{q^N+1}\ket{\mathbb{S}}.
\end{equation}
A two-copy average density matrix with statistical model representation $\ket H$ corresponds to an ensemble of quantum states that are indistinguishable from globally Haar-random states using second moment properties~\cite{dalzellRandomQuantumCircuits2022}---the states form a so-called spherical $2$-design. 

It is useful to notice that swapping the two copies of the density matrix is a symmetry of the noiseless evolution. This symmetry interchanges the two fixed points $\ket{\mathbb{I}}$ and $\ket{\mathbb{S}}$. Thus, this statistical model can be interpreted as a classical Ising system in a ferromagnetic phase \cite{vasseurEntanglementTransitionsHolographic2019, liEntanglementDynamicsNoisy2023}; this analogy will turn out to be a fruitful way to understand the physics of the circuit with noise.

\subsection{\label{sec:noisy} Review of noisy circuit results}

Having reviewed the noiseless circuit results, we now turn to the noisy case.
When noise is added ($\gamma > 0$), fidelity is no longer conserved and output density matrices always converge to the maximally mixed density matrix. 
If the error rate $\varepsilon = (1-q^{-2})\gamma$ of the single-qubit noise channel is sufficiently small,  Ref.~\cite{dalzellRandomQuantumCircuits2021a} showed that the global white noise approximation 
\begin{equation} \label{eq:whitenoise}
\rho \approx F \rho_{\text{ideal}} +  (1 - F) \frac{\mathbb{I}}{q^N}
\end{equation}
accurately predicts second-moment quantities such as fidelity and linear cross-entropy \cite{aruteQuantumSupremacyUsing2019}. 
As consequences, in the same regime of noise, XEB approximates fidelity, and both decay exponentially 
with a rate proportional to the noise rate and the total number of gates. This exponential decay takes the form 
\begin{equation} \label{eq:fiddecay}
F(d) \sim (1 - \varepsilon)^{Nd},
\end{equation}
where $\varepsilon$ is the error rate of the single-qudit noise channel and $d$ is the circuit depth. In fact, this form of the decay is universal for systems with single-qudit noise and two-qudit gate count $Nd/2$---it is independent of the circuit architecture or gate set \cite{dalzellRandomQuantumCircuits2021a}.

In the context of the statistical model, the global white noise regime is one in which the dynamics is approximately captured by a linear combination of the two fixed points of the noiseless transfer matrix:
\begin{equation}
\ket{\rho(d)} \approx a(d) \ket{\mathbb{S}} + (1 - a(d)) \ket{\mathbb{I}},
\end{equation}
where $a(d) = (q^{N}F - 1)/(q^{2N} - 1).$
While the maximally mixed density matrix continues to be a fixed point of $\T$ with non-zero noise, the other fixed point $\ket{\mathbb{S}}$ is converted into a metastable state. 
To lowest order in the noise, the decay of $\ket{\mathbb{S}}$ in one layer of the circuit is given by 
\begin{gather}
 \frac{\matrixel{\mathbb{S}}{\T}{\mathbb{S}}}{\braket{\mathbb{S}}} = \left(q^{-2} \gamma +  (1 - \gamma) \right)^N = (1 - \varepsilon)^N,
\end{gather}
where  $\varepsilon = \gamma (1 - q^{-2})$, and thus the prediction for the fidelity $F(d)$ at depth $d$ matches \cref{eq:fiddecay}.

Intuition for this regime can be gained in a `heralded noise' picture, which models the evolution with error events that occur at fixed locations \cite{dalzellRandomQuantumCircuits2021a}. If the noise rate is small enough, the noiseless evolution has sufficient time between error events to return the state to the fixed points $\ket{\mathbb{I}}$ and $\ket{\mathbb{S}}$. By expanding the full noisy evolution of $\T$ into a sum of terms corresponding to different locations and times of the errors, Ref.~\cite{dalzellRandomQuantumCircuits2021a} showed that a sufficient condition on the noise to be in this regime is 
\begin{equation}
\varepsilon \ll \frac{1}{N \ln N},
\end{equation}
while conjecturing that 
$
\varepsilon < \frac{c}{N}
$
for some constant $c$ would suffice.
Ref.~\cite{gaoLimitationsLinearCrossEntropy2021} also studied the statistical model with noise quantitatively for several values of $\varepsilon$ and several choices of gate set, showing that the global white noise approximation fails to describe XEB at larger noise values.

\section{Fidelity and XEB Transition}
\label{sec:phase transition} 

\begin{figure*}[!htb]
     \centering
     \includegraphics[width=\textwidth]{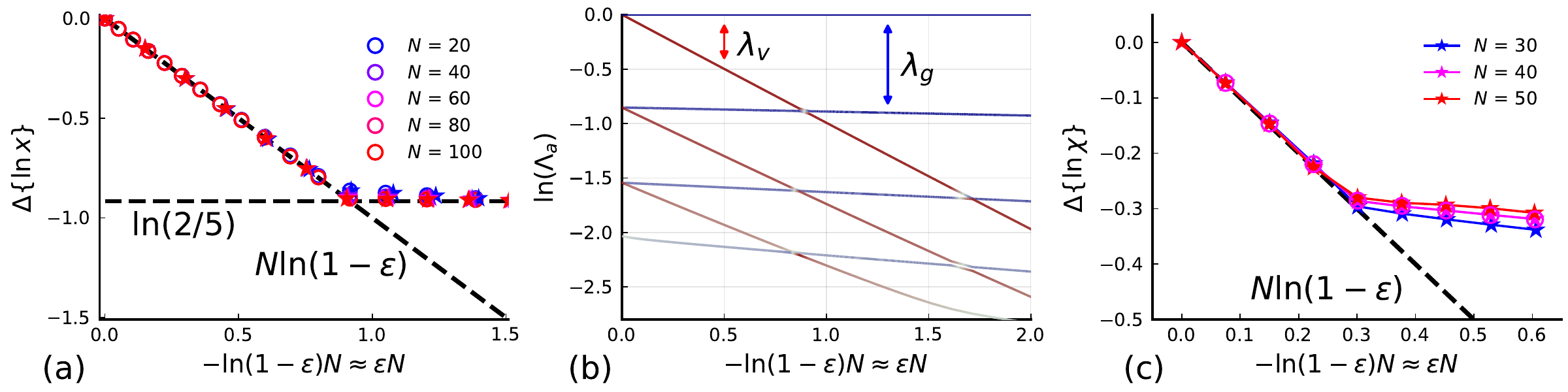}
    \caption{ \label{fig:transition}
         Decay rate $\Delta\{\ln \chi\}$ of XEB, when it has reached its asymptotic value (stars) and subleading eigenvalue $\Lambda_1$ of $\T$ (circles) for different values of the system size $N$, for the all-to-all circuit (a) and the 1D circuit (c), with Haar random gates.
         The decay rate of the XEB matches the subleading eigenvalue and undergoes a kink at the eigenvalue crossing. In (a), the kink occurs at $\varepsilon N = \log \frac{5}{2} \approx 0.92$.
        (b) Seven largest eigenvalues $\Lambda_a$ of the transfer matrix $\T$ in the all-to-all geometry with $N=40$ qubits and Haar random gates. States with low Hamming weight (blue tones) are insensitive to $\varepsilon N$, while states with extensive Hamming weight (red tones) have decay rates $\ln(\Lambda_a)$ linear in $\varepsilon N$.
    }
\end{figure*}

In this section, we will first look empirically at the behavior of fidelity and XEB beyond the global white noise approximation, and then explain the behavior with a spectral analysis of the transfer matrix $\T$. 
Our results confirm the conjecture of Ref.~\cite{dalzellRandomQuantumCircuits2021a}, showing that the global white noise approximation is indeed valid when $\varepsilon N < c$, with a sharp phase transition at $\varepsilon N = c$ separating the global white noise regime from a larger noise regime in which the metastable state $\ket{\mathbb{S}}$ decays too quickly to dominate the contributions to XEB. By contrast, the fidelity continues to track \cref{eq:fiddecay} in the larger noise regime $\varepsilon N > c$; thus, the phase transition also marks the point at which XEB fails to be a proxy for fidelity.

Figure \ref{fig:Fvsx} shows the behavior of average fidelity and XEB, computed for the all-to-all circuit geometry and Haar random two-qubit gates. The fidelity [Fig.~\ref{fig:Fvsx}(a)] is close to the global white-noise prediction \cref{eq:fiddecay} until saturation, when fidelity approaches a minimal value, $F = q^{-N}$, which is exactly the fidelity of any state with a maximally mixed state. Similar behavior occurs in one-dimensional circuits, as shown in \cref{app:oned} (\cref{fig:F-oned}).

By contrast, the decay rate of the XEB closely matches the global white noise value only for sufficiently small noise rates, as shown in Fig.~\ref{fig:Fvsx}(b). 
At higher noise, the decay rate of the XEB abruptly stops changing as noise is increased. To sharply distinguish the two regimes, one should look at the asymptotic decay rate of the XEB, 
\begin{equation}
\Delta \{\ln \chi(d)\} := \ln \frac{\chi(d+1)}{\chi(d)}.
\end{equation}
In all cases, $\chi(d+1)/\chi(d)$ reaches a plateau as a function of depth $d$, as shown in Fig.~\ref{fig:Fvsx}(c).
Empirically, we see two regimes: in the low-noise regime, $\chi(d+1)/\chi(d) \sim (1 - \varepsilon)^N$, while for higher noises $\chi(d+1)/\chi(d) \sim const.$, with a constant that is roughly independent of $N$ and $\varepsilon$.

\Cref{fig:transition} reveals the nature of the transition in the behavior of XEB. The asymptotic decay rate of XEB, shown in
\cref{fig:transition}(a) for the all-to-all circuit and in 
\cref{fig:transition}(c) for one-dimensional circuits, experiences a kink at a fixed value of $\varepsilon N$. Additionally, this rate is shown to precisely match the value of the largest eigenvalue $\Lambda_1$ of the transfer matrix, excluding the trivial largest eigenvalue $\Lambda_0 = 1$ corresponding to the maximally mixed density matrix fixed point. The kink in the value of $\Lambda_1$ occurs precisely at the location where the leading eigenvalues of $\T$ cross, as depicted in
\cref{fig:transition}(b) for the all-to-all model. 

This eigenvalue crossing is not accidental, but in fact a generic feature of a wide class of random circuit models.
The ubiquity can be explained via an analogy to the Ising model. At zero noise, the leading transfer matrix eigenvalues are doubly degenerate, corresponding to spontaneously symmetry broken states of the copy permutation symmetry; noise plays the role of a field that explicitly breaks the symmetry  \cite{liEntanglementDynamicsNoisy2023}. In 
\cref{fig:transition}(b), we see that this analogy fits as a description for the leading eigenvectors of $\T$, which can be split into two classes: eigenvectors with low Hamming weight, which are analogous to excitations on top of one ferromagnetic vacuum $\ket{\mathbb{I}}$, and eigenvectors with extensive Hamming weight, which correspond to excitations on top of the other ferromagnetic vacuum $\ket{\mathbb{S}}$.

Such an eigenvalue crossing would be visible not just in the dynamics of XEB, but in nearly all observables that can be computed by expectation values in the average two-copy density matrix. Suppose the eigenvalues of $\T$ are $\Lambda_a$, with $a$ being an index, with corresponding left-eigenvectors $\bra{v_a}$ and right-eigenvectors $\ket{v_a}$.
Then through the eigendecomposition of $\T$, any such observable will show dynamics of the form
\begin{equation} \label{eq:eigen}
    O(d) = \sum_a  c^O_a \Lambda_a^d, \text{ with }
    c^O_a = \frac{\braket{O }{ v_a} \braket{v_a }{ \rho_0}}{\braket{v_a }{ v_a}} ,
\end{equation}
where $\ket {\rho_0}$ is the initial product state and applying $\bra O$ corresponds to multiplying with $O$ and taking the trace. 
As we observe constant scaling gaps in the spectrum, we expect that this sum will be asymptotically dominated by the largest eigenvalue $\Lambda_a$ for which the corresponding coefficient $c^O_a$ is non-zero.

Having answered the question of why XEB experiences a transition in its decay rate, why do we not observe the same transition in the decay rate of the fidelity? To answer this, we analyze in detail the nature of the leading eigenvectors. 
Let us label the largest non-trivial eigenvalues $\Lambda_v$ and $\Lambda_g$ as depicted in \cref{fig:transition}(b),  
so that $\Lambda_1 = \max\{\Lambda_g, \Lambda_v\}$; additionally, we refer to the corresponding eigenvectors as $\ket{v_v}$ and $\ket{v_g}$, respectively. 

\begin{figure}[!h]
    \centering
    \includegraphics[width=\linewidth]{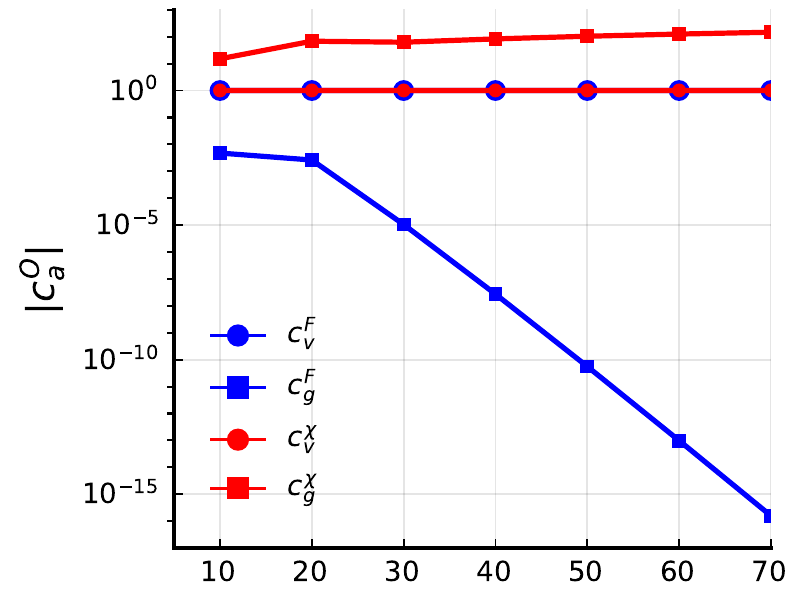}
    \caption{Coupling constants $c^O_a$ for fidelity and XEB with the leading non-trivial eigenvectors of the transfer matrix of the all-to-all model with Haar random gates and $\varepsilon N = 0.01$.}
    \label{fig:coupling}
\end{figure}

In the limit of large $N$ with fixed $\varepsilon N$, we find that the eigenvectors $\ket{v_v}$ and $\ket{v_g}$ are approximately constant, even as $\varepsilon N$ is tuned across the transition. We do not observe significant level repulsion or mixing of the eigenvectors at the crossing. This follows our expectations from the Ising analogy, as matrix elements between states that are oppositely polarized should be exponentially suppressed. 
In \cref{fig:coupling}, we show the behavior of the coefficients $c^O_a$ for the observables $F$ and $\chi$ in these eigenvectors. We see that while XEB couples significantly to each eigenvector,  $c^F_g$ approaches $0$ asymptotically as $q^{-N}$. This is a generic behavior, as $\ket{v_g}$ is an eigenvector with low Hamming weight but $\bra{\mathbb{S}}$, the observable vector corresponding to fidelity, is concentrated on strings with extensive Hamming weight.

Thus, the asymptotic behavior of fidelity can be described by truncating the sum in \cref{eq:eigen} to the three dominant eigenvalues, resulting in 
\begin{gather} \label{eq:threeeigenvectors}
   F \approx q^{-N} + (1 - \varepsilon)^{N d} +  \Lambda_g^d \frac {C}{q^N},
\end{gather}
for some constant $C$.
While the third term becomes larger than the second in the large noise phase as $d \to \infty$, the $q^{-N}$ suppression factor allows the second term to dominate the fidelity decay until the fidelity is close to saturation at $q^{-N}$ at depths $d \gtrsim (\ln q)/\varepsilon$.

In summary, we generically expect a transition in the decay rate of XEB but not fidelity at constant scaling depths. This transition occurs when $\Lambda_v = \Lambda_g$. 

\section{\label{sec:gatedependence} Location of the transition for different entangling gates}

\begin{figure}[t]
    \centering
    \includegraphics[width=\linewidth]{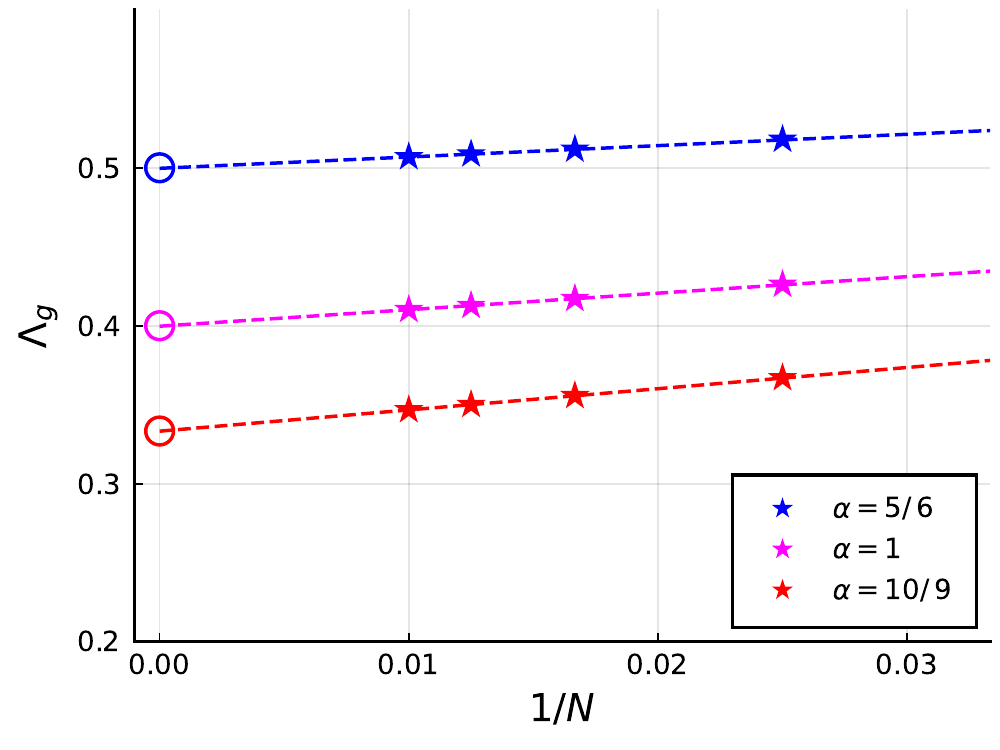}
    \caption{\raggedright Gap $\Lambda_g$ of the all-to-all circuit for several values of $\alpha$, computed for various system sizes $N$ as a function of $1/N$ (stars) and  linear fit (dashed lines). 
             In all cases, the extrapolation to infinite size via the linear fit matches the prediction $\Lambda_g = 1 - 3/5 \alpha$ (circles).
    \label{fig:all-to-all-gap} }   
\end{figure}

By the arguments of the previous section, the critical value corresponds to the location of the eigenvalue crossing $\Lambda_v((\varepsilon N)_c) = \Lambda_g((\varepsilon N)_c)$. 
In our numerics, these behave roughly as $\Lambda_g(\varepsilon N) = \Lambda_g(0) \equiv \Lambda_g$ and $\Lambda_v(\varepsilon N) = (1-\varepsilon)^N$. 
We expect this behavior of the eigenvalues to be exact in the thermodynamic limit based on the analogy to the Ising model explained in the previous section.
We confirm this expectation in the all-to-all architecture in \cref{fig:all-to-all-gap}, which shows that the finite-size corrections to the eigenvalues scale as $O(\varepsilon, 1/N)$, so that these corrections vanish as $N \rightarrow \infty$ keeping $\varepsilon N$ fixed.
Altogether, this gives us the criterion  
\begin{align} \label{eq:critpoint}
    (1 -\varepsilon)^N = \Lambda_g  \implies
    (\varepsilon N)_c \approx \lambda_g \equiv -\ln \Lambda_g,
\end{align}
for the critical value $(\varepsilon N)_c$ of $\varepsilon N$.

Thus, to solve for the critical value, it suffices to find the gap to the local excitation in the noiseless transfer matrix. In this section, we exploit this to locate the transition for various gate sets and geometries.

\begin{figure*}[t]
    \centering
     \begin{subfigure}[b]{0.48\textwidth}
         \centering
         \includegraphics[width=\textwidth]{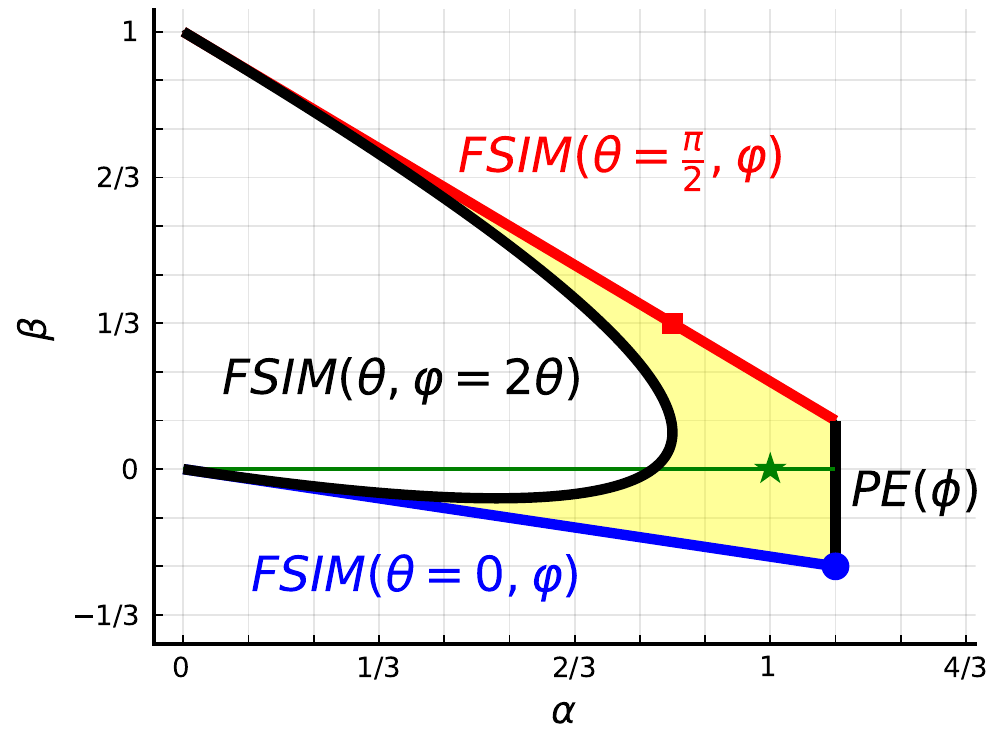}
         \caption{}
         \label{fig:region}
     \end{subfigure}
     \hfill
          \begin{subfigure}[b]{0.48\textwidth}
         \centering
         \includegraphics[width=\textwidth]{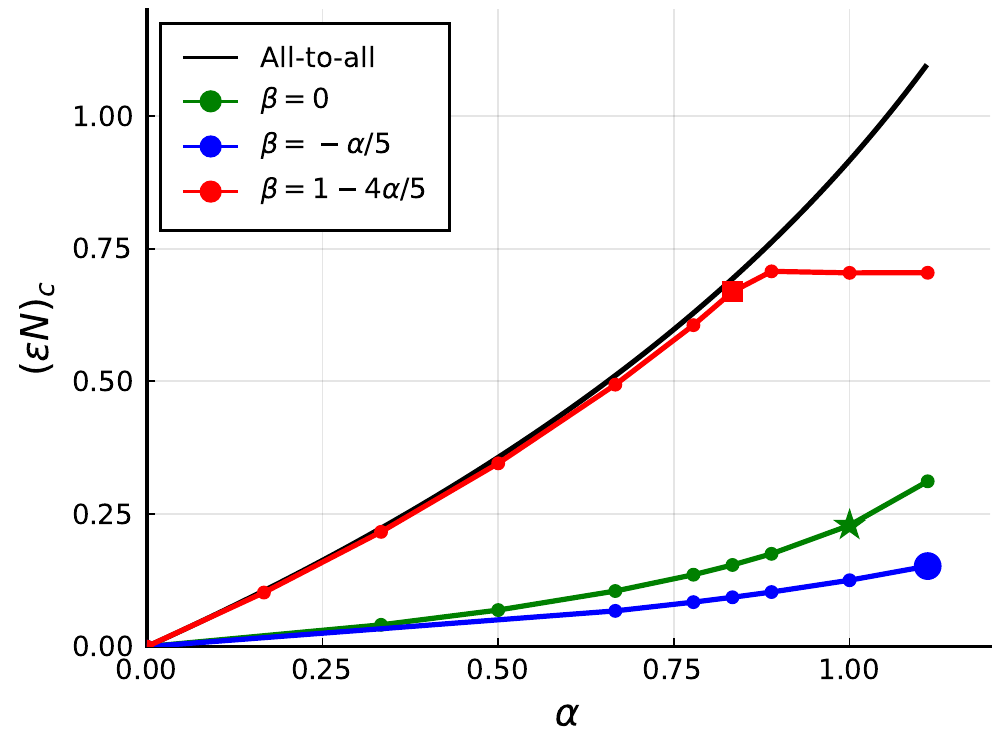}
         \caption{}
         \label{fig:crit}
     \end{subfigure}
    \caption{\raggedright (a) The valid region of $\alpha, \beta$ accessible to gate sets with a fixed two qubit gate (shaded yellow), with parameterizations of gates that realize the borders. The red square, green star, and blue circle respectively denote $\fsim(\pi/2, \pi/6)$ used in Ref.~\cite{aruteQuantumSupremacyUsing2019}, Haar random gates, and the control-$Z$ gate. (b) Critical value of $\varepsilon N$ for various geometries computed via \cref{eq:critpoint}. In the all-to-all geometry (black), we use the analytical formula \cref{eq:predictedgap}. For 1D circuits (green, blue, and red), we numerically compute $\Lambda_g$ along the three correspondingly colored lines in (a). }
    \label{fig:gatedependence}
\end{figure*}
Our discussion and results in the previous sections have focused on the case of Haar-random two-qudit gates, for which the statistical model update rule is given by \cref{eq:M} with $\alpha=1$ and $\beta=0$. 
However, in experimental settings, it is more convenient to use gate sets with a fixed choice of two-qubit gate~\cite{aruteQuantumSupremacyUsing2019,wu_strong_2021,zhu_quantum_2022}.
We consider in this section gate sets with a fixed two-qubit gate dressed with Haar-random single-qubit gates\footnote{
The statistical model for such gate sets was also derived in Ref.~\cite{gaoLimitationsLinearCrossEntropy2021}.}.
In \cref{app:deriv}, we derive the values of $\alpha$ and $\beta$ for general two-qubit gates ($q=2$); it is shown that $\alpha$ and $\beta$ are restricted to the region shown in \cref{fig:region}. We also derive explicit parameterizations for the gates that correspond to the boundaries of this region.
Three of the boundaries are given by specific parameterizations of the so-called fSim gate \cite{foxen_demonstrating_2020},
\begin{align}
    \fsim(\theta,\varphi) = \begin{pmatrix}
        1 & 0 & 0 & 0 \\
        0 & \cos \theta & - i  \sin \theta & 0 \\
        0 & - i \sin \theta & \cos \theta & 0 \\
         0 & 0 & 0 & e^{- i \varphi}
    \end{pmatrix}, 
\end{align}
while the last boundary corresponds to so-called \emph{perfect entanglers},
\begin{align}
    \pe(\phi) =
\begin{pmatrix}
     \sin \phi & 0 & 0 & -i \cos \phi \\
     0 & \cos \phi  & -i \sin \phi & 0 \\
      0 & -i \sin \phi & \cos \phi & 0  \\ 
     -i \cos \phi & 0 & 0 &  \sin \phi
 \end{pmatrix}.
\end{align}

As discussed in \cref{sec:all}, the transfer matrix for permutation-symmetric circuit architectures has no dependence on $\beta$. Thus, for the all-to-all circuit, we only need to compute $\Lambda_g(\alpha)$. Empirically, we find that $\Lambda_g(\alpha) = 1 - 3\alpha/5$, using finite-size diagonalizations for various $\alpha$ and $N$ and extrapolating to $N\to\infty$. This is shown in \cref{fig:all-to-all-gap}.

We now present a heuristic argument that in the all-to-all model the exact formula for the gap of the transfer matrix is 
\begin{equation} \label{eq:predictedgap}
    \Lambda_g = (1 - \alpha) + \alpha \frac{2}{q^2+1},
\end{equation}
which matches our empirical finding with $q=2$.
Consider configurations $\sigma$ on $N$ sites with a small fraction $f$ of $\SWAP$ factors in random locations, that is, configurations $\sigma$ with low Hamming weight. 
When the density of such sites is small, each individual gate encounters either zero or one such site, as the chance of a collision is $\mathcal{O}(f^2)$. 
At each gate that encounters a $\SWAP$, there is a chance of $1-\alpha$ that the action of the gate leaves one $\SWAP$, and a chance of  $\alpha/(q^2+1)$ that it produces two $\SWAP$ factors; thus in total the expected $\SWAP$ density will be decreased from $f$ to $((1 - \alpha) + \alpha \frac{2}{q^2+1}) f$. 
States of the form 
\begin{equation}
\ket{\rho} \propto  \ket{\mathbb{I}} + a \ket{v_g} 
\end{equation}
satisfy our hypothesis of having only a small expected fraction $f$ of $\SWAP$ factors, and after an application of the transfer matrix have a fraction $\Lambda_g f$ of $\SWAP$ factors. Thus it seems reasonable to expect that \cref{eq:predictedgap} holds.

The situation becomes more intricate in low-dimensional architectures in which the critical point depends on both $\alpha$ and $\beta$. 
To analyze the dependence on the values of $\alpha$ and $\beta$, we numerically compute---and show in \cref{fig:crit}---the critical value $-\ln \Lambda_g$ in the one-dimensional brickwork architecture [Fig.\ \ref{fig:cartoon}(b)] along the upper and lower boundaries of the allowed region shown in \cref{fig:region}, as well as the line $(\alpha,\beta=0)$. The computations use the MPS-Krylov algorithm discussed in \cref{sec:1d}, with $N=40$ sites and $\varepsilon=0$.
We find a monotonic dependence of the critical point on both $\alpha$ and $\beta$. In particular, the dependence of the critical value of the  $\fsim(\pi/2, \varphi)$-gate on $\alpha$ roughly matches that of the all-to-all architecture until it saturates at $\Lambda_g = 1/2$, corresponding to $(\varepsilon N)_c = \ln 2$, while the smaller values of $\beta$ have a smaller critical value.\footnote{Let us also note that, in agreement with Ref.~\cite{gaoLimitationsLinearCrossEntropy2021}, we find that the noise robustness of the $\fsim(\pi/2, \pi/6)$ gate from Ref.~\cite{aruteQuantumSupremacyUsing2019} exceeds that of Haar-random gates, which in turn exceeds that of controlled-$Z$ gates.
}
We leave it as an interesting question to precisely understand this saturation effect at maximal values of $\beta$. 
\medskip

Our results show that the critical value of $\varepsilon N$ can be tuned by varying $\alpha $ and $\beta$. 
In the all-to-all architecture, it can be tuned all the way from $0$ to a maximal value of $\ln 3 \approx 1.1$  by tuning $\alpha$ from $0$ to the maximal value $10/9$. 
In the one-dimensional architecture, it can be tuned from $0$ to $\ln 2 \approx 0.69$ at the maximal value of $\alpha$ and $\beta$. The corresponding gate is the iSWAP gate \cite{gaoLimitationsLinearCrossEntropy2021}. 
We expect our results to qualitatively carry over to any architecture.

The entangling gate parameter $\alpha$ is related to a property of the gate known as its \emph{entangling power} \cite{zanardi_entangling_2000,wang_entangling_2003}. 
Heuristically, at every value of $\alpha$, the gate parameter $\beta$  measures the \emph{swapping power} of the corresponding gate---for the minimal value of $\beta$ given by $-\alpha/5$, the gate does not swap the qubits at all, while at its maximal value $1- 4\alpha/5$ it corresponds to a full swap. 
Indeed, multiplying the gates on the blue line (minimum swapping power) with a SWAP gate gives rise to the red line (maximum swapping power); in general multiplying a gate with entangling and swapping power $\alpha$ and $\beta$ with a SWAP gate, gives rise to a gate with  $\alpha' = \alpha$, $\beta' = 1- \alpha - \beta$.
Our results show that computational benchmarking experiments should pick maximally entangling and maximally swapping gates if they wish to maximize the noise level at which the phase transition occurs.

\section{\label{sec:con} Outlook}
We have given strong analytic and numerical evidence for the existence of a phase transition in the dynamics of noisy random circuits as a function of noise.
Below the phase transition, the white noise model of Ref.~\cite{dalzellRandomQuantumCircuits2021a} suffices to describe second moment quantities of the density matrix.
In particular, in this regime, sample-efficient proxies for the fidelity such as the linear cross-entropy benchmark mimic the fidelity well.
Above the phase transition, this correspondence fails to hold.
Indeed, in this regime, the XEB and fidelity behave asymptotically differently. 
Even more drastically, the decay rate of the XEB fails to match that of the fidelity ($F$) or its shifted version ($F-q^{-n}$).
Our results hold for any Haar-invariant two-qudit gate set and any single-qudit noise channel (except leakage error). However, as the qualitative results only rely on features that are universal to Ising-like systems with explicit symmetry breaking fields, we expect that they generalize to systems with arbitrary locally-acting and Haar-invariant gate sets and noise channels.

Our results imply that near-term quantum experiments purporting to use the linear cross-entropy benchmark as a proxy for fidelity must pay careful attention to the location of this phase transition.
Our results are shown here for the all-to-all geometry and the one-dimensional geometry---we expect the physics of the phase transition to be similar in other experimentally relevant geometries, such as two dimensions, as well.
The location of the phase transition depends both on the geometry and the properties of the gate set used.
The transfer matrix approach outlined here may be used in order to estimate the location of the transition for any gate set and geometry.

A separate question that we have not dealt with in this work is the ability to spoof the XEB metric, i.e., the ability to efficiently produce samples that achieve a high XEB score without necessarily producing samples from a state close in fidelity.
The noisy output state above the phase transition indeed achieves an XEB score that is higher than the fidelity, but it is unclear whether this output distribution may be classically efficiently simulated.
Recent work \cite{Aharonov2022} shows that this is possible for depths logarithmic in system size and larger.
Conversely, \textcite{gaoLimitationsLinearCrossEntropy2021} show that, by adversarially selecting which circuit locations experience strong noise, ``high'' values of the XEB can be achieved classically. 
It remains an exciting question to what extent the phase transition we have identified in the XEB as a function of noise strength is reflected in the computational complexity of classically simulating the corresponding noisy circuit.

\textit{Note added.}---After completing this work, we became aware of a complementary work studying the same phase transition \cite{morvan23}.

\section{\label{sec:ack} Acknowledgments}
We thank M.~Endres, S.~Gopalakrishnan, H.~Pichler, D.~Huse, X.~Gao, J.~Choi, and C.D.~White for helpful discussions. 
This research was performed while B.W. held an NRC Research Associateship award at the National Institute of Standards and Technology.
M.J.G.~and P.N.~acknowledge support from the NSF QLCI (award No.~OMA-2120757) and NSF grant PHY-1748958 through the KITP program on ``Quantum Many-Body Dynamics and Noisy Intermediate-Scale Quantum Systems,'' where the idea for this work was conceived.
A.V.G.~and P.N.~were supported in part by the DoE Quantum Systems Accelerator, DoE ASCR Accelerated Research in Quantum Computing program (award No.~DE-SC0020312),  NSF QLCI (award No.~OMA-2120757), DoE ASCR Quantum Testbed Pathfinder program (award No.~DE-SC0019040), NSF PFCQC program, AFOSR, ARO MURI, AFOSR MURI, and DARPA SAVaNT ADVENT. 
D.H.\ acknowledges financial support from the US Department of Defense through a QuICS Hartree fellowship.
This research was also supported in part by the National Science Foundation under Grant No. NSF PHY-1748958.
\bibliography{main,references}

\appendix

\section{\label{app:deriv} Derivations of statistical model rules}
In this Appendix, we rederive the rules governing the statistical model following the notation of Ref.~\cite{dalzellRandomQuantumCircuits2021a}.
First, we derive the rule corresponding to the application of an arbitrary two-qudit gate followed by an average over single-qudit gates.
Next, we derive the rule corresponding to the application of single-qudit \com{stochastic?} noise.

The initial state $(\ketbra{0})^{\otimes N}$ is first acted upon by single-qudit Haar-random gates, which project the two-copy state into the symmetric subspace spanned by $\{I,\SWAP\}^{\otimes N}$.
As in the main text, $I$ is the ($q^2 \times q^2$) $2$-copy identity and $\SWAP$ the SWAP operation of two qudits acting between the two copies.
Upon further applications of arbitrary two-qudit gates followed by Haar-random single-qudit gates, the state remains in the $\{I,\SWAP\}^{\otimes N}$ space.

We now examine the effect of an arbitrary two-qubit gate $U$.
The update rule is obtained by performing the following average over Haar-random single-qubit gates $V$ and $W$ acting on the two qubits in the two copies of the system:
\begin{align}
\mathbb E_{V,W}\left[
\begin{tikzcd}[row sep={0.8cm,between origins},transparent]
\qw & \gate{V} & \gate[2]{U} & \gate[4]{\rhotwo} & \gate[2]{U^\dag} & \gate{V^\dag} &\qw\\
\qw & \gate{W} & \qw & & &\gate{W^\dag} &\qw\\
\qw & \gate{V} & \gate[2]{U} & &\gate[2]{U^\dag} & \gate{V^\dag} &\qw \\
\qw & \gate{W} & \qw & & & \gate{W^\dag} &\qw 
\end{tikzcd}\right]. \label{eq:gateupdates}
\end{align}

We parameterize the two-qubit unitary as
\begin{align}
\label{eq:canonical gate}
U = 
\begin{pmatrix}
 \cos(c_-) & 0 & 0 & -i \sin(c_-)  \\
 0 & \cos(c_+)e^{ic_3} &-i \sin(c_+) e^{i c_3} & 0 \\
 0 & -i \sin(c_+) e^{i c_3} & \cos(c_+)e^{ic_3} & 0 \\
 -i \sin(c_-) & 0 & 0 & \cos(c_-) \\
\end{pmatrix},
\end{align}
where $c_\pm = \frac{c_1 \pm c_2}{2}$ with $c_1, c_2 \in \mathbb R$.
Up to local single-qubit rotations, this is the most general two-qubit unitary \cite{Makhlin2002,Zhang2003}.
Since single-qubit gates are averaged over, all gates that are equivalent up to local single-qubit gates yield the same update rules.
The invariants of such two-qubit gates under local rotations have been characterized by Makhlin \cite{Makhlin2002}.
Given a gate of the form \eqref{eq:canonical gate}, these are given by
\begin{align}
G_1 &= \frac{e^{-2ic_3}}{4}(\cos(c_1-c_2) +e^{2ic_3}\cos(c_1+c_2))^2,\\
G_2 &= \cos 2c_1 + \cos 2c_2 + \cos 2c_3, \label{eq:generalgateinvariants}
\end{align}
and they can be efficiently computed for any gate \cite{Makhlin2002}. 
The quantity $\abs{G_1} = \frac{1}{4}(1+ \cos 2c_1 \cos 2c_2 + \cos 2c_2 \cos 2c_3 + \cos 2c_3 \cos 2c_1)$ is related to the entangling power of the two-qubit gate.
The relation of the entangling power of a gate to a specific property of the transfer matrix was noted by \textcite{gaoLimitationsLinearCrossEntropy2021}.

The two-copy state $\rhotwo$ is a combination of the operators\footnote{To ease readability, in the following, we abbreviate $A \otimes B = A B $ for $A, B \in \{I, \SWAP\}$.} $II, I\SWAP, \SWAP I,$ and $\SWAP \SWAP$ , where the SWAP operation is between the two copies, i.e., between either the first and third qubits (in the case of $\SWAP I$) or the second and fourth qubits (in the case of $I \SWAP$) in \cref{eq:gateupdates}.
The operators $II$ and $\SWAP \SWAP$ are left invariant under the action of the gates in \cref{eq:gateupdates}.
It suffices to consider $I\SWAP$, as the evolution of $\SWAP I$ may be derived from that of $I\SWAP$ by exchanging the two copies.
Denote by $\overline{I \SWAP}$ the evolution of $I \SWAP$ under the update of a single gate layer \cref{eq:gateupdates}.
Expanding $\overline{I \SWAP} = a II + b I\SWAP + c\SWAP I + d\SWAP \SWAP$ in the basis $\{I, \SWAP\}$, the coefficients $a,b,c,$ and $d$ are determined through the Hilbert-Schmidt inner product with the basis operators.
Specifically, we have
\begin{align}
\Tr [(\overline{I \SWAP}) (II) ]&= 2^4 a + 2^3b + 2^3 c + 2^2 d = 2^3, \label{eq:overlap1}\\
\Tr [(\overline{I \SWAP})( I\SWAP )]&= 2^3 a + 2^4 b + 2^2 c + 2^3 d = x, \\
\Tr [(\overline{I \SWAP})( \SWAP I)] &= 2^3 a + 2^2  b + 2^4c + 2^3 d = y, \\
\Tr [(\overline{I \SWAP})( \SWAP \SWAP)] &= 2^2 a + 2^3 b + 2^3 c + 2^4d = 2^3, \label{eq:overlap4}
\end{align}
with $x=6+4\abs{G_1} + 2G_2$ and $y = 6 + 4\abs{G_1} - 2G_2$.
The prefactors can be easily found from the fact that $\tr[I] =  2^2$, while $\tr[IS] = 2$. Moreover, the constraints \eqref{eq:overlap1} and \eqref{eq:overlap4} are determined by the facts that $\tr[(\overline{I \SWAP})(II)] = \tr[(I \SWAP)(II)]$ since the trace is conserved and $\tr[(\overline{I \SWAP})(\SWAP \SWAP)] = \tr[(I \SWAP)(\SWAP \SWAP)]$ since the fidelity is conserved.
Solving these equations gives us
\begin{align}
a = d = \frac{20-x-y}{18}, \\
b = \frac{4x+y-32}{36}, \\
c = \frac{x+4y-32}{36}.
\end{align}
The transfer matrix in terms of the normalized operators $\{I/q^2, \SWAP/q\}$ is then
\begin{align}
 M &=
 \begin{pmatrix}
    1 & 2a & 2a & 0 \\
    0 & b & c & 0 \\
    0 & c & b & 0 \\
    0 & a/2 & a/2 & 1
 \end{pmatrix},
\end{align}
implying $\alpha = \frac{5a}{2} = \frac{10(1-\abs{G_1})}{9}$ and $\beta = c = \frac{-1}{18} - \frac{G_2}{6} + \frac{5}{9}\abs{G_1}$ from the comparison with \cref{eq:M}.
Plugging in the easily-computed values of $G_1$ and $G_2$ for any gate, we obtain the transfer matrix
\begin{align}M=
\begin{pmatrix}
1 & \frac{8}{9}(1-\abs{G_1}) & \frac{8}{9}(1-\abs{G_1}) & 0 \\
0 & \frac{-1}{18} + \frac{G_2}{6} +\frac{5}{9}\abs{G_1} & \frac{-1}{18} - \frac{G_2}{6} + \frac{5}{9}\abs{G_1} & 0 \\
0 & \frac{-1}{18} - \frac{G_2}{6} + \frac{5}{9}\abs{G_1} & -\frac{1}{18} + \frac{G_2}{6} +\frac{5}{9}\abs{G_1} & 0 \\
0 & \frac{2}{9}(1-\abs{G_1}) & \frac{2}{9}(1-\abs{G_1}) & 1
\end{pmatrix}.
\end{align}

\subsection*{Notation of Gao et al.}
We briefly take a detour to compare our notation with the notation of \textcite{gaoLimitationsLinearCrossEntropy2021}.
The basis of their statistical mechanical model $\{I,\Omega\}$ is related to our basis of $\{I/q^2, \SWAP/q\}$ through the linear transformation effected by the matrix $O$:
\begin{align}
O &= \begin{pmatrix}
 1 & 1 \\
 0 & q^2 -1
\end{pmatrix}, \text{with} \\
O^{-1} &= \begin{pmatrix}
 1 & \frac{-1}{q^2-1} \\
 0 & \frac{1}{q^2-1}
\end{pmatrix}.
\end{align}
Performing a similarity transform of transfer matrix $\T$ using $O\otimes O$ yields a transfer matrix in the basis $\{I,\Omega\}$
\begin{align}
\begin{pmatrix}
1 & 0 & 0 & 0 \\
0 & 1- \beta - \frac{\alpha q^2}{q^2+1}  & \beta + \frac{\alpha}{q^2+1}& \frac{\alpha}{q^2+1} \\
0 & \beta + \frac{\alpha}{q^2+1} & 1- \beta - \frac{\alpha q^2}{q^2+1} & \frac{\alpha}{q^2+1}\\
0 & \frac{\alpha (q^2-1)}{q^2+1} & \frac{\alpha (q^2-1)}{q^2+1}& 1-\frac{2\alpha}{q^2+1}
\end{pmatrix},
\end{align}
implying
\begin{align}
D &= \frac{\alpha q^2}{q^2+1} + \beta, \\
R &= \frac{\alpha (q^2-1)}{q^2+1}, \\
\eta &= q^2-1, 
\end{align}
in the notation of \citeauthor{gaoLimitationsLinearCrossEntropy2021}.

\subsection*{Transfer matrix parameters for some gates}
We will use the parametrization with $\alpha$ and $\beta$ from the main text in order to discuss the possible transfer matrices associated with various gates.
The various gates and their parameter values are given in \cref{tab:gateparameters}.

\begin{table*}[t]
\scriptsize
\renewcommand{\arraystretch}{1.4}
\begin{tabular}{@{}l@{\hspace{0.5em}}l@{\hspace{1em}}l@{\hspace{0em}}l@{\hspace{1em}}l@{\hspace{1em}}l@{\hspace{1em}}l@{}} \toprule
\textbf{Gate}
& \multicolumn{2}{c}{\textbf{Gate invariants}} & \multicolumn{2}{c}{ \bf $\{I/q^2, \SWAP/q\}$ basis} & \multicolumn{2}{c}{\bf  $\{I, \Omega\}$ basis} \\ \cmidrule(lr){2-3} \cmidrule(lr){4-5} \cmidrule(lr){6-7}
& $|G_1|$ & $G_2$ & $\alpha$ & $\beta$ & $R$ & $D$ \\ \midrule
Controlled-$U$ & $G_1$ &2$G_1$+1 & $\frac{10}{9}(1-\abs{G_1})$ & $\frac{-1}{18} - \frac{G_2}{6} + \frac{5}{9}\abs{G_1}$ & $\frac{2}{3}(1-\abs{G_1})$ & $\frac{5}{6} -\frac{G_2}{6} - \frac{1}{3}\abs{G_1}$  \\
CNOT & $0$ & $1$ & $\frac{10}{9}$ & $\frac{-2}{9}$ & $\frac{2}{3}$ & $\frac{2}{3}$\\
SWAP & $1$ & $-3$ & $0$ & $1$ & $0$ & $1$ \\
iSWAP & $0$ & $-1$ & $\frac{10}{9}$ & $\frac{1}{9}$ & $\frac{2}{3}$ & $1$ \\
Haar-random gate & -- & -- & $1$ & $0$ & $\frac{3}{5}$ & $\frac{4}{5}$ \\ 
Single-qubit gate & $1$ & $3$ & $0$ & $0$ & $0$ & $0$ \\ \midrule
fSIM($\theta,\varphi$) & \parbox{6.5em}{\raggedright $\frac{1}{4}(1+\cos^2 2\theta +2\cos 2\theta \cos \varphi)$} & \parbox{6em}{\raggedright $2\cos 2\theta +\cos \varphi$} & \parbox{6.5em}{\raggedright $\frac{5}{36}(5-\cos 4\theta -4\cos 2\theta \cos \varphi)$} & \parbox{12em}{\raggedright$\frac{1}{72}(11 + 5\cos 4\theta -24\cos 2\theta + 20 \cos 2\theta \cos \varphi - 12 \cos \varphi)$} & \parbox{6.5em}{\raggedright$\frac{1}{12}(5-\cos 4\theta -4\cos 2\theta \cos \varphi)$} & \parbox{12em}{\raggedright$\frac{1}{24}(17 - \cos 4\theta -8\cos 2\theta - 4 \cos 2\theta \cos \varphi - 4 \cos \varphi)$} \\
fSIM($\theta=\frac{\pi}{2},\varphi$) & $\frac{1}{2}(1-\cos \varphi)$ & $-2+\cos\varphi$ & $\frac{5}{9}(1 + \cos \varphi)$ & $\frac{1}{9}(5 - 4\cos \varphi)$ & $\frac{1}{3}(1 + \cos \varphi)$ & $1$ \\
fSIM($\theta=0,\varphi$) & $\frac{1}{2}(1+\cos \varphi)$ & $2+\cos \varphi$ & $\frac{5}{9}(1 - \cos \varphi)$ & $\frac{-1}{9}(1 - \cos \varphi)$ & $\frac{1}{3}(1- \cos \varphi)$ & $\frac{1}{3}(1- \cos \varphi)$ \\ \midrule
\parbox[][][l]{9em}{\raggedright Special perfect entanglers: \\
$c_1=\pi/2, c_2=2\phi - \frac{\pi}{2}, c_3=0$} & 0 & $-\cos 4\phi$ & $\frac{10}{9}$ & $\frac{1}{18}(-1+3\cos 4\phi)$ & $\frac{2}{3}$ & $\frac{1}{6}(5+\cos 4\phi)$
\\ \bottomrule
\end{tabular}
\caption{\raggedright Gate invariants and transfer matrix parameters for some common gates.
}
\label{tab:gateparameters}
\end{table*}

We note the constraints on the possible values that $G_1$ and $G_2$, and hence the transfer matrix parameters, can take.
First, we use the relation between $\abs{G_1}$ and the entangling power of the gate \cite{Balakrishnan2010}:
\begin{align}
    e_p(U) = \frac{2}{9}(1-\abs{G_1}).
\end{align}
From this and from the fact that the maximum entangling power is between 0 and 2/9, we have the constraints $0\leq \abs{G_1} \leq 1$, giving $0\leq \alpha \leq 10/9$, or $0\leq R \leq 2/3$.
The gates that maximize entangling power have been characterized \cite{Balakrishnan2010,Zhang2003}.
These take the form $c_1 = \pi/2, c_2 = 2\phi - {\pi}/{2}, c_3=0$, and we denote the gate $\pe(\phi)$, defined in the main text.

Reference~\cite{gaoLimitationsLinearCrossEntropy2021} had also noted that $0\leq D \leq 1$.
In addition, they proved that $D\geq R$, which has an interesting interpretation in terms of the entangling power of a gate characterized by $R$ and the apparent entangling power characterized by $D$.
In terms of the quantities $\cos 2c_1 \cos 2c_2 + \cos 2c_2 \cos 2c_3 + \cos 2c_3 \cos 2c_1$ and $\cos 2c_1 + \cos 2c_2 + \cos 2c_3$ that characterize any gate, these two inequalities translate to 
\begin{align}
 \cos 2c_1 \cos 2c_2 + \cos 2c_2 \cos 2c_3 + \cos 2c_3 &\cos 2c_1 + \nonumber \\ 
 \frac{1}{2}(\cos 2c_1 + \cos 2c_2 + \cos 2c_3) & \geq \frac{-3}{2}, \\
 \cos 2c_1 \cos 2c_2 + \cos 2c_2 \cos 2c_3 + \cos 2c_3 &\cos 2c_1 \nonumber - \\
  \frac{1}{2}(\cos 2c_1 + \cos 2c_2 + \cos 2c_3) & \leq \frac{3}{2}.
\end{align}
These also imply $\beta + \alpha/(q^2+1) \geq 0$ and $\beta \leq 1- \alpha q^2/(q^2+1)$.
It may be seen that the $\fsim$ gate with $\theta=0$ and arbitrary $\varphi$ saturates the $\beta \geq -\alpha/5$ constraint, with $D = R = (1-\cos\varphi)/3$.
Similarly, it can also be seen that the inequality $\beta \leq 1-4\alpha/5$ is saturated with the $\fsim$ gate with $\theta=\pi/2$ and arbitrary $\varphi$, with the values $\alpha = 5(1+\cos\varphi)/9$ and $\beta = (5-4\cos\varphi)/9$.

We find a new constraint using the inequality $(x+y+z)^2 \geq 3(xy+yz+zx)$ for any $x,y,z \in \mathbb{R}$.
Applying this to $x= \cos 2c_1, y = \cos 2c_2, z = \cos 2c_3$, we obtain
\begin{align}
 G_2^2 - 3(4\abs{G_1}-1) &\geq 0, \text{or } \\
 D-R &\leq \left(D-\frac{R}{2} \right)^2, \text{or } \\
 \beta + \frac{\alpha}{5} &\leq \left(\beta + \frac{\alpha}{2}\right)^2.
\end{align}
As shown in \cref{fig:region}, these constraints account for the entire boundary of the region of parameters accessible via a single two-qubit unitary.
Furthermore, the equality is satisfied when $c_1=c_2=c_3$.
It may be checked that the fSIM gate with parameters $(\theta, \varphi=2\theta)$ also satisfies these conditions and has
\begin{align}
 \alpha &= \frac{5}{12}(1-\cos 4\theta), \\
 \beta &= \frac{1}{72}(21+15\cos 4\theta -36\cos 2\theta).
\end{align}

As an interesting aside, we also point out that a \emph{fixed} two-qubit gate leads to the same update rules as a Haar-random two-qubit gates.
This can be found by solving for $c_1, c_2$, and $c_3$ in our parametrization \cref{eq:generalgateinvariants}: there are several solutions, but one of them is given by
\begin{align}
 c_1 = \frac{\pi}{4}, c_2 = \frac{-1}{2}\tan^{-1}\sqrt{\frac{2}{3}}, c_3 = \frac{\pi}{2} + c_2.
\end{align}

\subsection*{Statistical model rules for noise channels}

Similarly to the update for unitary gates, the update for the noise channel can be obtained by performing an average over a single Haar-random single-qudit gate $V$ in the two copies of the system. There are two distinct rules, depending on whether the noise acts on one or both copies of the system. If noise acts on one copy, then the necessary average $\mathcal{N}_1(\rhotwo)$ takes the form 
\begin{align}
\sum_a \mathbb{E}_V\left[
\begin{tikzcd}[row sep={0.8cm,between origins},transparent]
\qw & \gate{V} & \gate[1]{\mathcal{K}_a} & \gate[2]{\rhotwo} & \gate[1]{\mathcal{K}_a^{\dagger}} & \gate{V^\dag} &\qw\\
\qw & \gate{V} & \qw & & \qw & \gate{V^\dag} &\qw
\end{tikzcd}
\right],
\label{eq:onecopynoisediagram}
\end{align}
where $\mathcal{K}_a$ are the Kraus operators associated with the noise channel $\mathcal{N}$.
If noise acts on both copies, then the average $\mathcal{N}_2(\rhotwo)$ instead takes the form 
\begin{align}
\sum_{a,b} \mathbb{E}_V\left[
\begin{tikzcd}[row sep={0.8cm,between origins},transparent]
\qw & \gate{V} & \gate[1]{\mathcal{K}_a} & \gate[2]{\rhotwo} & \gate[1]{\mathcal{K}_a^{\dagger}} & \gate{V^\dag} &\qw\\
\qw & \gate{V} & \gate[1]{\mathcal{K}_b} & & \gate[1]{\mathcal{K}_b^{\dagger}} & \gate{V^\dag} &\qw
\end{tikzcd}\right].
\label{eq:twocopynoisediagram}
\end{align}

We begin with the expression for a single-qubit Haar average of an operator $O$:
\begin{align}
 \mathbb{E}_V \left[ (V\otimes V) O (V\otimes V)^\dag \right] =: \overline{O} = a I + b \SWAP, \text{ where} \nonumber\\
 a = \frac{\Tr (O) - \frac{1}{q}\Tr (O\SWAP)}{q^2-1}, \quad b = \frac{\Tr (O\SWAP) - \frac{1}{q}\Tr (O)}{q^2-1}. \label{eq:haaravgformula}
\end{align}
Since the Kraus operators are trace-preserving, we have
$\Tr\left( \mathcal{N}_1(O)\right) = \Tr\left( \mathcal{N}_2(O)\right)= \Tr(O)$.
The only parameters that govern the noise dynamics are then $\Tr\left( \mathcal{N}_1(O)\cdot \SWAP\right)$ and $\Tr\left( \mathcal{N}_2(O)\cdot \SWAP\right)$ for $O\in\{I,\SWAP\}$.

Consider $\Tr\left( \mathcal{N}_1(I)\cdot \SWAP\right)$.
After noise acts on the state, the operator becomes
\begin{align}
\mathcal{N}_1(I/q^2) = \mathbb{E}_V\left[
\begin{tikzcd}[row sep={0.8cm,between origins},transparent]
\qw & \gate{V} & \gate[1]{\rho_1}  & \gate{V^\dag} &\qw\\
\qw & \gate{V} &  \gate[1]{I/q}&  \gate{V^\dag} &\qw
\end{tikzcd}\right],
\end{align}
where $\rho_1$ is the state the identity $I/q$ maps to under the nonunital noise.
Therefore $\Tr\left( \mathcal{N}_1(I/q^2)\cdot \SWAP\right) = \Tr \left( \rho_1 \cdot I/q \right) = \Tr \rho_1/q = 1/q$.
Thus, this parameter is also independent of the noise strength and type, giving us three remaining parameters.

\begin{figure*}
    \centering
    \includegraphics[width=\textwidth]{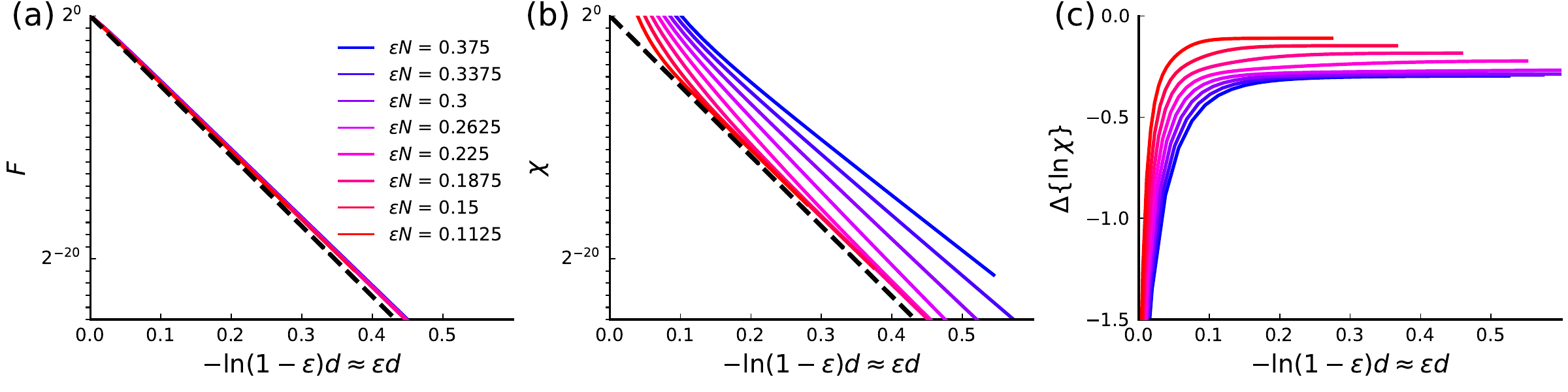}
       \caption{
       Fidelity and XEB in 1D circuits with $N=40$ qubits and Haar-random gates, for various values of the noise rate $\varepsilon$ (see color code in (a)), plotted against rescaled circuit depth.
   (a) Fidelity $F$ curves collapse onto the global-white noise prediction $(1-\varepsilon)^{Nd}$ until saturating at $q^{-N}$.
   (b) Unnormalized XEB $\chi$. For low noise $\varepsilon N \lesssim 0.22$, the XEB decay rate matches that of fidelity. (c) Decay rate of the XEB. For $\varepsilon N \gtrsim 0.22$, the decay rate saturates to a $\varepsilon N$-independent constant.
   \label{fig:Fvsx-oned}}
\end{figure*}

Let us denote these as 
\begin{align}
 \Tr\left( \mathcal{N}_2(I)\cdot \SWAP\right) &\eqqcolon q+\mu, \label{eq:tracevalues_i2} \\
 \Tr\left( \mathcal{N}_1(\SWAP)\cdot \SWAP\right) &\eqqcolon Y_1, \label{eq:tracevalues_s1}\\
\Tr\left( \mathcal{N}_2(\SWAP)\cdot \SWAP\right) &\eqqcolon Y_2. \label{eq:tracevalues_s2}
\end{align}
In terms of these parameters, from \cref{eq:haaravgformula,eq:tracevalues_i2,eq:tracevalues_s1,eq:tracevalues_s2}, we have
\begin{align}
\mathcal{N}_2 \left[\frac{I}{q^2}\right] &= \left(1 - \frac{\mu}{q(q^2-1)}\right) \frac{I}{q^2} + \frac{\mu}{q(q^2-1)} \frac{\SWAP}{q},\\
\mathcal{N}_1 \left[\frac{\SWAP}{q}\right] &= \frac{q^2-Y_1}{q^2-1} \frac{I}{q^2} + \frac{Y_1 - 1}{q^2-1} \frac{\SWAP}{q},\\
\mathcal{N}_2 \left[\frac{\SWAP}{q}\right] &= \frac{q^2-Y_2}{q^2-1} \frac{I}{q^2} + \frac{Y_2 - 1}{q^2-1} \frac{\SWAP}{q}.
\end{align}

The parameters are chosen so that in the case of unital noise, we have $\mu = 0$.
The interpretation of $\mu$ is that $\Tr\left( \mathcal{N}_2(I)\cdot \SWAP\right) = q^2 \Tr [\rho_1^2] = q+\mu$.
Therefore $\mu = q^2 \Tr [\rho_1^2] - q$, which measures how impure the state $\rho_1$ is.
We call $\mu$ the \emph{nonunitality}.

\textcite{dalzellRandomQuantumCircuits2021a} have derived the relation between $Y_1$ and $Y_2$ and the noise channel's infidelity and nonunitarity for unital noise channels.
We follow their derivation and obtain generalized expressions for nonunital noise as well.
Before coming to the evolution of the SWAP operator, let us relate the channel's infidelity and nonunitarity to the quantities $Y_1$ and $Y_2$.
From Ref.~\cite{dalzellRandomQuantumCircuits2021a}, the average infidelity is given by
\begin{align}
r &= 1- \mathbb{E}_V \Tr \left[V \ketbra{\psi}V^\dag \mathcal{N}_1\left(V \ketbra{\psi}V^\dag\right) \right] \\
&= 1- \Tr \left(\SWAP \mathcal{N}_1 \left( \frac{I+\SWAP}{q(q+1)} \right) \right) \\
 &= 1 - \frac{1}{(q^2-1)(q+1)}\left(q(q^2-1)+Y_1 \frac{q^2-1}{q}\right) \\
 &= \frac{q-Y_1/q}{q+1}, \text{ implying} \\
 Y_1&= q^2-rq(q+1).
\end{align}
Similarly, the unitarity is given by
\begin{align}
 u &= \frac{q}{q-1}\left( \mathbb{E}_V \Tr \left[(\mathcal{N}_2(V\ketbra{\psi}V^\dag))^2 \right] - \frac{1}{q}\right) \\
 &= \frac{q}{q-1} \Tr \left[\SWAP \mathcal{N}_2 \left( \frac{I+\SWAP}{q(q+1)} \right) \right] - \frac{1}{q-1} \\
 &= \frac{Y_2 + \mu -1}{q^2-1}, \text{ implying} \\
 Y_2&= (q^2-1)u + 1- \mu.
\end{align}

Thus, the transfer matrices are as follows:
\begin{equation}
N^{(1)}(\N) = 
    \begin{bmatrix}
        1  & \gamma_1 \\
        0 & 1-\gamma_1
    \end{bmatrix},
\end{equation}
where
$\gamma_1 = {qr}/{(q-1)}$.

With two-copy noise, we have instead
\begin{equation}
N^{(2)}(\N) = 
    \begin{bmatrix}
        1 - \delta_2  & \gamma_2 \\
        \delta_2 & 1-\gamma_2
    \end{bmatrix},
\end{equation}
where 
\begin{align}
 \delta_2 &= \frac{\mu }{q(q^2-1)}, \\
 \gamma_2 &= 1-u + \frac{\mu}{q^2-1}. 
\end{align}

\section{\label{sec:purity} Purity and collision probabilities for noisy circuits}

In this Appendix, we consider second-moment properties of the density matrix in which both copies experience noise. As such, we no longer have the liberty to only consider the infidelity $\gamma$ of the noise channel $\N$.

The averaged purity of the noisy evolution has the same form as fidelity except both copies of the circuit are evolved with the noisy evolution:
\begin{align}
    P(\N) &= \Eu \Tr (\rhoN\rhoN) \nonumber \\
    &= \Eu  \Tr(\SWAP^{\otimes N}  \rhoN \otimes \rhoN).
    \label{eq:def-P}
\end{align}

Similarly, the collision probability of the
noisy circuit has the same form as the linear cross entropy benchmark except both copies are evolved with the noisy evolution:
\begin{align}
    Z(\N) &= \mathbb{E}_{\mathbf{U}}\sum_\x \bra{\x}\rhoN \ket{\x} \bra{\x} \rhoN \ket{\x} \nonumber \\
    &= \Eu  \Tr(\PROJ^{\otimes N} \rhoN \otimes \rhoN).
    \label{eq:def-Z2}
\end{align}

When both copies of the density matrix experience noise, the update rules for noise take the form of a Markov
matrix \(N^{(2)}\):
\begin{equation}
N^{(2)} = 
    \begin{bmatrix}
        1 - \delta_2  & \gamma_2 \\
        \delta_2 & 1-\gamma_2
    \end{bmatrix}.
\end{equation}
Derivations of formulas for $\gamma_2$ and $\delta_2$ are shown in Appendix~\ref{app:deriv}.

For unital noise, $\delta_2 = 0$, and thus $N^{(2)}$ has the same form as $N(\gamma)$ considered throughout the paper. In this case, the update rules for computing purity and the collision probability are identical to those of computing fidelity and the linear XEB, but with $\gamma_2$ in place of $\gamma$:
\begin{align}
P(\gamma_2, \delta_2=0) &= F(\gamma=\gamma_2),\\
Z(\gamma_2, \delta_2=0) &= X(\gamma=\gamma_2).
\end{align}
Thus, the phase transition described above for XEB can be seen additionally in the collision probability of noisy circuits with unital noise. 

However, realistic digital quantum devices experience non-unital noise. Non-zero $\delta_2$ dramatically changes the physics of the statistical model, as the maximally mixed density matrix is no longer the equilibrium state reached at large depths. We leave further considerations of non-unital noise to future work.\\

\section{Additional numerical data} \label{app:oned}

In \cref{fig:Fvsx-oned}, we illustrate that, like the all-to-all circuits, one dimensional circuits show fidelity that decays like $(1 - \varepsilon)^{Nd}$ for nearly all depths before saturation, while the XEB decay rate goes through a transition. The corresponding decay rate of XEB is shown in 
\cref{fig:transition}(c) to match the leading eigenvalue $\Lambda_1$ of the transfer matrix and to experience a kink at the eigenvalue crossing, just as in the all-to-all circuit. Thus, we see that the same phenomenology that describes the all-to-all circuit appears in the 1D circuit.

\begin{figure}[h]
     \centering
     \includegraphics[width=\linewidth]{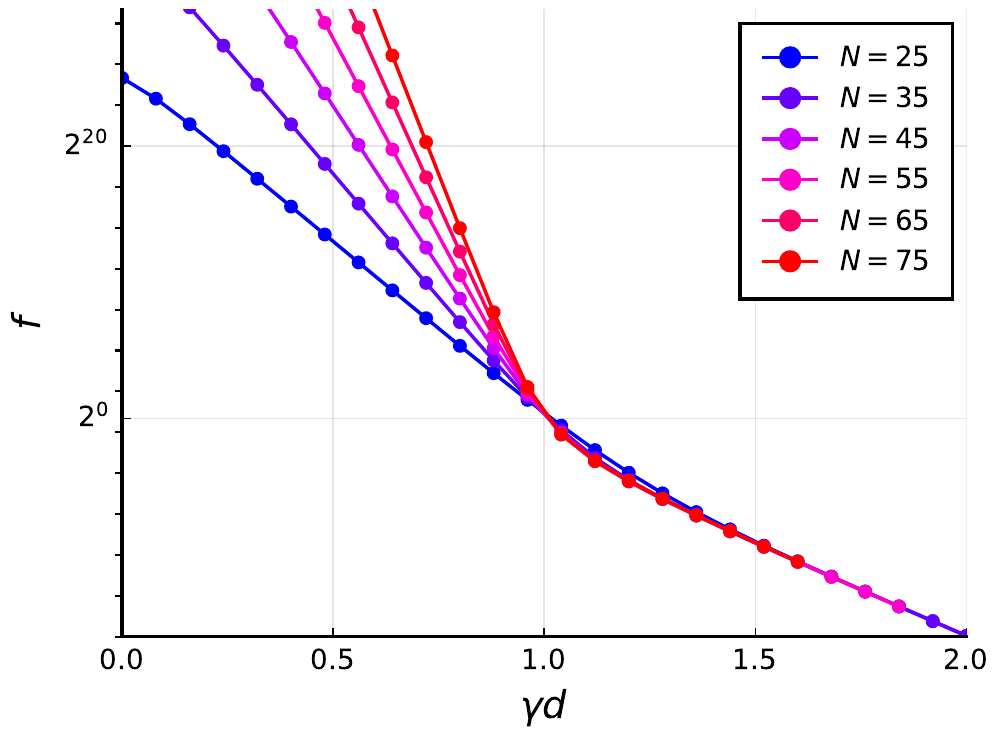}
     \caption{Shifted fidelity in one-dimensional circuits with Haar random gates and $\varepsilon=0.03$, for various system sizes $N$, plotted versus rescaled depth $\gamma d = \frac43 \varepsilon d$.}
     \label{fig:shiftedf-oned}
 \end{figure}

While the fidelity decays like $(1 - \varepsilon)^N$ until saturation at $q^{-N}$, we can additionally study the shifted-and-rescaled fidelity $f = q^{N} F - 1$. By subtracting off the saturation value, we see in \cref{fig:shiftedf-oned}, at sufficiently large depths, the $(1 - \varepsilon)^N$ decay is dominated by an $N$-independent decay. This confirms that the coupling constants between fidelity and locally-gapped eigenvectors is non-zero. 

\end{document}